\documentclass[aps,prd,amsmath,amssymb,eqsecnum,nofootinbib]{revtex4}

\allowdisplaybreaks[1]
\usepackage{graphicx}
\usepackage{verbatim}
\usepackage{color}
\newcommand{\xxp}{\left( x,x' \right)}
\begin{document}

\title{Quasi-local contribution to the scalar self-force: Geodesic Motion}
\author{Adrian C. Ottewill}
\email{adrian.ottewill@ucd.ie}
\author{Barry Wardell}
\email{barry.wardell@ucd.ie}
\affiliation{School of Mathematical Sciences, University College Dublin, Belfield, Dublin 4, Ireland}

\date{\today}

\begin{abstract}
We consider a scalar charge travelling in a curved background spacetime. We calculate the quasi-local contribution to the scalar self-force experienced by such a particle following a geodesic in a general spacetime. We also show that if we assume a massless field and a vacuum background spacetime, the expression for the self-force simplifies significantly. We consider some specific cases whose gravitational analog are of immediate physical interest for the calculation of radiation reaction corrected orbits of binary black hole systems. These systems are expected to be detectable by the LISA space based gravitational wave observatory. We also investigate how alternate techniques may be employed in some specific cases and use these as a check on our own results.
\end{abstract}
\maketitle

\section{Introduction}
\label{sec:intro}
There has been considerable recent interest in the modelling of astronomical and cosmological gravitational wave sources. This interest has been spurred on by the recent and upcoming construction of both ground and space based gravitational wave observatories. Ground based detectors such as LIGO \cite{LIGO}, VIRGO \cite{VIRGO} and GEO600 \cite{GEO600} have already reached the data collection phase.\
The primary space based gravitational wave observatory, LISA \cite{LISA} - the Laser Interferometer Space Antenna - is still in the design and planning phase. A joint NASA/ESA mission, it is expected to launch within the next few years. It will detect gravitational waves in the frequency band 0.1-0.0001Hz, a band in which a considerable number of interesting gravitational wave sources are expected to be found.

In order to extract the maximum amount of information from these gravitational wave detectors, data analysis techniques such as matched filtering may be employed. To improve the effectiveness of this data analysis, it is necessary to have accurate predictions of the waveforms of the gravitational radiation emitted by candidate sources. One of the primary candidate sources of such a gravitational wave signal falls under the category of large mass binary systems. In particular, extreme mass ratio systems such as that of a compact solar mass object (mass $m$) inspiralling into a black hole of $10^3$ to $10^8$ solar masses (mass $M$) are expected to be detectable by LISA out to Gpc distances.

As a prerequisite to the calculation of the waveform templates for such sources, it is necessary to model the orbital evolution of the system. In the case of extreme mass ratio systems, we can do so by making the approximation that the smaller mass is travelling in the background spacetime of the larger mass. However, the compact nature of the smaller mass means that it itself will distort the curvature of the spacetime in which it is moving to a non-negligible extent. As a result, it does not follow a geodesic of the background spacetime of the larger mass, but rather, it follows a geodesic of the total spacetime of both masses. In this sense the motion of the smaller mass is affected by the presence of its own mass, i.e. the smaller mass is seen to exert a force on itself. This is called the self-force. Clearly, the computation of this self-force is of fundamental importance to the accurate calculation of the orbital evolution of such binary systems and hence to the prediction of the gravitational radiation waveform.

Because of this non-linear behavior, it is extremely difficult, if not impossible, to exactly solve the equations of motion of such a system. However, due to the extreme mass ratio involved, the deviation of the object's trajectory from the background geodesic is small over sufficiently small time-scales (less than the natural length-scale of the background). In this case, to linear order in $m$, we can model the deviation from the background as a spin-2 field generated by the particle and living on the background spacetime. This field is then seen to couple to the smaller object to cause the self-force.

In the present work, we will focus not on the gravitational self-force described thus far, but rather on the analogous but simpler scalar self-force. Instead of considering the smaller mass, $m$, to be generating a gravitational field, we take it to have a scalar charge, $q$. This charge couples to a massless scalar field which is then the cause of the self-force. Otherwise, we leave the problem exactly as posed above. In making this change, the resulting analysis remains extremely similar to the gravitational case, but the algebraic details of the calculations become simpler. This will allow us to explore the problem and develop techniques which we will at a later date apply to the physically more interesting gravitational case.

An expression for the electromagnetic self-force was originally obtained by DeWitt and Brehme \cite{DeWitt:1960fc} (later corrected by Hobbs \cite{Hobbs}) and more recently recovered by Quinn and Wald \cite{Quinn:1996am}. In addition, Ref. \cite{Quinn:1996am} derives an expression for the gravitational self-force, as was done previously by Mino, Sasaki and Tanaka \cite{Mino:1996nk}. Using a generalization of his axiomatic approach, Quinn also obtained an expression for the scalar self-force \cite{Quinn:2000wa}. These results show how the self-force may be expressed in terms of an integral of the retarded Green's function over the entire past world-line of the particle. This retarded Green's function possesses two terms, one representing geodesic propagation, the other a tail-term arising from back-scattering by the spacetime geometry. (Solutions to wave equations on most curved manifolds depend not only on Cauchy data directly intersecting the past light cone, but also on Cauchy data interior to that intersection. The portion of the field that propagates in the null directions is called the \emph{direct} part, while the portion of the field that propagates in the timelike directions inside the light cone is called the \emph{tail}.) Clearly, the tail part of a particle's field can interact with the particle, leading to a contribution to the self-force. 

Although in principle this formal solution gives the desired result, in practice the calculation of the Green's function poses a formidable challenge. There has been a significant amount of work in this area in recent years. Various techniques have been employed. There has been considerable recent success using mode-sum approaches, amongst others \cite{Detweiler:2002gi,Haas:2007kz,Barack:2007tm,Barack:2007jh,Barack:2007we}. Other work has focused on a matched expansion approach and also on obtaining exact solutions under specific circumstances. For the Schwarzschild background, Wiseman \cite{Wiseman:2000rm} found an exact solution for the scalar self-force on a particle held at rest. Anderson, Hu and Eftekharzadeh \cite{Anderson:2005ds} built on this work along with other results to derive an expression for the self-force on a particle held at rest, then released and allowed to fall radially inward. Anderson and Wiseman \cite{Anderson:2005gb} also examined the matched expansion approach under the two cases of static particles and circular geodesic paths in a Schwarzschild background. For the case of a gravitational field, Anderson, Flanagan and Ottewill \cite{Anderson:2004eg} have evaluated the quasi-local contribution to the self-force up to $O \left( \Delta \tau ^3 \right)$. Comprehensive reviews of the radiation-reaction and self-force problem are given by Poisson \cite{Poisson:2003nc} and by Detweiler \cite{Detweiler:2005kq}.

In this paper, we use the results of D\'{e}canini and Folacci \cite{Decanini:2005gt} as a basis for our own work. In section \ref{sec:recapQuinn} we first recap the work of Quinn \cite{Quinn:2000wa} to define what we mean by the quasi-local part of the self-force and then we derive an expression for it and for the equations of motion in terms of the Green's function. Next, in section \ref{sec:cov} we use this result to calculate an expression for the quasi-local portion of the scalar self-force in terms of the expansion coefficients given by Ref. \cite{Decanini:2005gt}, and in terms of the proper time to the matching point $\Delta \tau$ and the 4-velocity of the particle, $u^\alpha$.

By making the assumption of motion in a massless field on a vacuum background spacetime, we show in section \ref{sec:simplify} that the resulting self-force expression simplifies considerably and in fact vanishes up to and including third order in $\Delta \tau$.

In section \ref{sec:cases}, we examine some particularly interesting and simple cases in both Schwarzschild and Kerr spacetimes. We then show that using a corrected form of the results of Ref. \cite{Anderson:2005gb}, we obtain identical results, order by order for these special cases. This confirms the consistency of our results and  the validity of our approach. We also present corrected results for the higher order terms in Ref. \cite{Anderson:2005gb} and give updated truncation error graphs.

Throughout this paper, we use units in which $G=c=1$ and adopt the sign conventions of \cite{Misner:1974qy}. We denote symmetrization of indices using brackets (e.g. $(\alpha \beta)$) and exclude indices from symmetrization by surrounding them by vertical bars (e.g. $(\alpha | \beta | \gamma)$). Roman letters are used for free indices and Greek letters for indices summed over all spacetime dimensions.

\section{The Quasi-local Scalar Self-force and Equations of Motion}
\label{sec:recapQuinn}
Consider a particle of scalar charge $q$ and mass $m$ travelling in a curved background spacetime. We treat this as a point particle and will refer to it as a particle throughout this paper. However, the arguments presented here are equally valid for non-point objects provided the object's extended structure does not affect its center-of-mass motion.

The scalar field generated by the particle propagates along null geodesics, while the particle itself will (approximately) follow a time-like geodesic of the background. The field then interacts with the particle, effectively causing a force, called the scalar self-force. This self-force may be expressed as a sum of local and non-local parts \cite{Quinn:2000wa}
\begin{equation}
\label{eq:FullForce}
f^{a} = q^2 \left( \frac{1}{3} \left( \dot{a}^{a} - a^2 u^{a} \right) + \frac{1}{6} \left( R^{a \beta} u_{\beta} + R_{\beta\gamma} u^\beta u^\gamma u^a \right) +\left(\frac{1}{2}m_{\rm field}^2 - \frac{1}{12} \left( 1 - 6 \xi \right) R \right) u^a \right) + \lim _{\epsilon \rightarrow 0} q^2 \int_{-\infty}^{\tau - \epsilon} \nabla^{a} G_{ret} \xxp d\tau '
\end{equation}
where $u^{\alpha}$ is the 4-velocity of the particle, $a$ is the 4-acceleration, $\dot{a} = \frac{\partial{a}}{\partial{\tau}}$ is the derivative of the 4-acceleration with respect to proper time and $G_{ret}$ is the retarded scalar Green's function. Here, we have trivially extended the expression given in Ref. \cite{Quinn:2000wa} and Ref. \cite{Poisson:2003nc} to include terms corresponding to the field mass $m_{\rm field}$ and coupling to the background scalar curvature, $\xi$.

In this form, the local terms are all given explicitly, so their calculation can be carried out immediately without posing any real difficulty. Our task is therefore to elucidate the non-local integral term. To this end, we will work with an expression for the self-force which just contains the non-local integral term:
\begin{equation}
f^{a} = \lim _{\epsilon \rightarrow 0} q^2 \int_{-\infty}^{\tau - \epsilon} \nabla^{a} G_{ret} \xxp d\tau '
\end{equation}
We do so with the understanding that the local terms can easily be added back in later if necessary.

This leaves us with an expression for the scalar self-force in terms of an integral of the gradient of the retarded Green's function over the entire past world line of the particle. The limiting feature at the upper limit of integration is necessary because of the singular nature of the Green's function at the point $x=x'$.

This integral would normally be evaluated using a mode-sum approach. One would calculate a sufficient number of the modes and sum them up to obtain the Green's function solution and hence the self-force. However, the limiting feature in the upper limit of integration causes great difficulties for this method. As $\epsilon$ decreases, it is necessary to take an increasing number of modes to obtain an accurate solution. When $\epsilon$ tends towards 0, the number of modes required becomes infinite and it is no longer possible to accurately compute the integral. However, we can avoid this problem if we split the integral in two parts
\begin{equation}
\label{eq:SplitForce}
f^{a} = \lim _{\epsilon \rightarrow 0} q^2 \int_{\tau - \Delta \tau}^{\tau - \epsilon} \nabla^{a} G_{ret} \xxp d\tau ' + q^2 \int_{-\infty}^{\tau - \Delta \tau} \nabla^{a} G_{ret} \xxp d\tau '
\end{equation}
where $\tau - \Delta \tau$ is a \emph{matching point}. It is chosen so that both integrals may be evaluated separately using different means and then combined to obtain an expression for the total self-force. The second integral in this expression may be evaluated, for example, by using the methods of quantum field theory in curved spacetime \cite{Casals:2005kr} to perform a mode-sum calculation of the Green's function away from the singularity $x=x'$.  We call the first term in Eq. (\ref{eq:SplitForce}) the \emph{quasi-local} part of the self-force. It is the computation of this integral that is the main focus of the remainder of this paper.

The greatest obstacle to obtaining a solution for this quasi-local integral lies with the difficulty in the computation of the retarded Green's function, $G_{ret}$. It is a solution of the wave equation on a curved background spacetime:
\begin{equation}
\label{eq:Wave}
\left( \Box _{x} - m^2 - \xi R \right) G_{ret} \xxp = - 4\pi \delta \xxp
\end{equation}
In general, it proves extremely difficult, if not impossible, to find an exact solution of Eq. (\ref{eq:Wave}). However, provided $x$ and $x'$ are sufficiently close, we can use the Hadamard form for the retarded Green's function solution \cite{Quinn:2000wa},
\begin{equation}
\label{eq:Hadamard}
G_{ret}\xxp = \theta_{-} \xxp \left\lbrace U \xxp \delta \left( \sigma \xxp \right) - V \xxp \theta \left( - \sigma \xxp \right) \right\rbrace 
\end{equation}
where $\theta_{-} \xxp$ is analogous to the Heaviside step-function (i.e. $1$ when $x'$ is in the causal past of $x$, $0$ otherwise), $\delta \xxp$ is the standard Dirac delta function, $U \xxp$ and $V \xxp$ are symmetric bi-scalars having the benefit that they are regular for $x' \rightarrow x$ and $\sigma \xxp$ is the Synge \cite{Synge,Poisson:2003nc,DeWitt:1965jb} world function. For the time-like geodesics that are of interest to this paper, $\sigma$ is equal to minus one half of the squared geodesic distance between $x$ and $x'$:
\begin{equation}
\sigma \xxp = -\frac{1}{2} s^2
\end{equation}

This form of the retarded Green's function is valid provided $x'$ lies within a normal neighborhood of $x$, i.e. provided there is a unique geodesic connecting $x$ and $x'$. Clearly, by simply choosing an appropriate value for $\Delta \tau$, we can satisfy this normal neighborhood condition and use the Hadamard form for the retarded Green's function in the first integral in Eq. (\ref{eq:SplitForce}).\footnote{Anderson and Wiseman \cite{Anderson:2005gb} have calculated some explicit values for the normal neighborhood boundary for a particle following a circular geodesic in Schwarzschild spacetime. These values place an upper bound estimation on the geodesic distance in the past at which we can place the matching point $\Delta \tau$.} 

The first term (involving $U \xxp$) in Eq. (\ref{eq:Hadamard}) is called the \emph{direct} part of the Green's function and does not contribute to the self-force. This can be understood by the fact that the presence of $\delta \left( \sigma \xxp \right)$ in the direct part of $G_{ret}$ means that it is only non-zero on the past light-cone of the particle. However, since the quasi-local integral in Eq. (\ref{eq:SplitForce}) is totally internal to the past light-cone, this term does not contribute to the integral.

The second term in Eq. (\ref{eq:Hadamard}) is known as the \emph{tail} part of the Green's function. It is this term that is responsible for the self-force and its calculation is the primary focus of this paper.

We can now substitute Eq. (\ref{eq:Hadamard}) into the quasi-local part of Eq. (\ref{eq:SplitForce}) to obtain the final form of the quasi-local self-force:
\begin{equation}
\label{eq:QLSFInt}
f^{a}_{\rm QL} = -q^2 \int_{\tau - \Delta \tau}^{\tau} \nabla^{a} V \xxp d\tau '
\end{equation}
Note that it is no longer necessary to take the limit since $V\xxp$ is regular everywhere.

Unlike the cases of electromagnetic and gravitational fields, where the 4-acceleration is exactly equal to the self-force divided by the mass, there is one further step required in order to obtain the equations of motion in the presence of a scalar field. In this case, the self-force contains two components, one related to the 4-acceleration, the other related to the change in mass:

\begin{equation}
f^{a}_{\rm QL} = \frac{dm}{d\tau} u^{a} + m a^{a}
\end{equation}

The 4-acceleration is always orthogonal to the 4-velocity, so in order to obtain it from the self-force, we take the projection of $f^{a}$ orthogonal to $u^{a}$:

\begin{equation}
\label{eq:ma}
ma^{a} = P^{a}_{\beta} f^{\beta}
\end{equation}

where
\begin{equation}
\label{eq:Projection}
P^{b}_{a} = \delta ^{b}_{a} + u_{a} u^{b}
\end{equation}
is the projection orthogonal to $u^\alpha$.

The remaining component of the self-force (which is tangent to the 4-velocity) is simply the change in mass:

\begin{equation}
\label{eq:dmdtau}
\frac{dm}{d\tau} = -f^{\alpha} u_{\alpha}
\end{equation}

\section{Covariant Calculation}
\label{sec:cov}
Within the Hadamard approach, the symmetric bi-scalar $V\xxp$ is expressed in terms of an expansion in increasing powers of $\sigma$ \cite{Decanini:2005gt}:
\begin{equation}
\label{eq:V}
V \xxp = \sum_{n=0}^{\infty} V_{n}\xxp \sigma ^{n}\xxp
\end{equation}
Substituting the expansion (\ref{eq:V}) into Eq.~(\ref{eq:Wave}), using the identity $\sigma_{;\alpha} \sigma^{;\alpha} = 2 \sigma$ and explicitly setting the coefficient of each manifest power of $\sigma^{n}$ equal to zero, we find (in 4-dimensional space-time) that the coefficients $V_{n}\xxp$ satisfy the recursion relations
\begin{equation}
\label{eq:RecursionV}
\left( n+1 \right) \left( 2n +4 \right) V_{n+1} + 2 \left( n+1 \right) V_{n+1;\mu}\sigma ^{;\mu} - 2 \left( n+1 \right)V_{n+1}\Delta ^{-1/2}\Delta ^{1/2}_{\phantom{1/2} ;\mu} \sigma^{;\mu} + \left( \Box _{x} - m^2 - \xi R \right) V_n = 0 \quad \mathrm{for}~ n \in \mathbb{N}
\end{equation}
along with the initial conditions
\begin{subequations}
\label{eq:InitialV}
\begin{eqnarray}
2V_0 + 2V_{0;\mu}\sigma ^{;\mu} - 2V_0 \Delta ^{-1/2}\Delta ^{1/2}_{\phantom{1/2} ;\mu} \sigma^{;\mu} + \left( \Box _{x} - m^2 - \xi R \right) U_0 &=& 0\\
U_0 &=& \Delta ^{1/2}
\end{eqnarray}
\end{subequations}
The quantity $\Delta \xxp$ is the Van Vleck-Morette determinant defined as \cite{Decanini:2005gt}
\begin{equation}
\Delta \xxp = - \left[ -g \left( x \right) \right] ^{-1/2} det \left( -\sigma _{;\mu \nu '} \xxp \right) \left[ -g \left( x' \right) \right] ^{-1/2}
\end{equation}

It is important to note that the Hadamard expansion is not a Taylor series and is not unique. Indeed, in Appendix~\ref{sec:DeWittVacuum} we show how, under the assumption of a massless field on a vacuum background spacetime, the terms from $V_0$ and $V_1$ conspire to cancel to third order when combined to form $V$. Such cancellation is also seen, for example, in deSitter space-time, where for a conformally invariant theory all the $V_n$'s are non-zero while $V\equiv0$. 

In order to solve the recursion relations (\ref{eq:RecursionV}) and (\ref{eq:InitialV}) for the coefficients $V_{n} \xxp$, it is convenient to first write $V_n \xxp$ in terms of its covariant Taylor series expansion about $x = x'$:
\begin{equation}
\label{eq:Vn}
V_n \xxp = \sum _{p=0}^{\infty} \frac{ \left( -1 \right) ^p }{p!} v_{n (p)} \xxp
\end{equation}
where the $v_{n (p)}$ are bi-scalars of the form
\begin{equation}
v_{n (p)} \xxp = v_{n\,\,\alpha_1 \dots \alpha_p} (x) \sigma ^{;\alpha_1} \xxp \dots \sigma ^ {;\alpha_p} \xxp
\end{equation}

D\'{e}canini and Folacci \cite{Decanini:2005gt} calculated expressions for these $v_{n (p)}$ up to $O \left( \sigma ^{2} \right)$, or equivalently $O \left( \Delta \tau ^{4} \right)$, where $\Delta \tau$ is the magnitude of the proper time separation between $x$ and $x'$. In Appendix \ref{sec:cov-5}, we use the symmetry of the Green's function in $x$ and $x'$ to extend their results by one order in the proper time separation, $\Delta \tau$, or equivalently one half order in $\sigma \xxp$.

We now turn to the evaluation of the self-force. We could calculate the self-force given this form for $V_n \xxp$, however, it proves easier to work with $V\xxp$  expressed in the form of a covariant Taylor expansion, i.e. an expansion in increasing powers of $\sigma^{;a}$:

\begin{equation}
\label{eq:Vt}
V \xxp = \sum _{p=0}^{\infty} \frac{ \left( -1 \right) ^p }{p!} v_{\alpha_1 \dots \alpha_p} (x) \sigma ^{;\alpha_1} \xxp \dots \sigma ^ {;\alpha_p} \xxp
\end{equation}

Explicit expressions for the $v_{a_1 \dots a_p} (x)$,  up to $O \left( \sigma^{5/2} \right)$, can then be easily obtained order by order by combining the relevant $v_{n\,\,a_1 \dots a_p}$ as described in Appendix \ref{sec:DeWittVacuum}.

Substituting Eq. (\ref{eq:Vt}) into Eq. (\ref{eq:QLSFInt}), taking the covariant derivative and noting that the $v_{a_1 \dots a_p}$ are symmetric under exchange of all indices, $a_1, \dots, a_p$, we obtain an expression for the self-force in terms of $v_{a_1 \dots a_p}$, $\sigma$ and their derivatives:
\begin{eqnarray}
\label{eq:SigmaExpandedIntegral}
f^{a}_{\rm QL} &=& -q^2 \int_{\tau - \Delta \tau}^{\tau} \Big[
	v^{; a}
	- v^{\phantom{\beta} ; a}_{\beta} \sigma^{; \beta}
	- v_{\beta} \sigma^{; \beta a}
	+ \frac{2}{2!} v_{\beta \gamma} \sigma^{; \beta a} \sigma^{; \gamma}
	+ \frac{1}{2} v^{\phantom{\beta \gamma} ; a}_{\beta \gamma} \sigma^{; \beta} \sigma^{; \gamma}
	- \frac{3}{3!} v_{\beta \gamma \delta} \sigma^{; \beta a} \sigma^{; \gamma} \sigma^{; \delta}
	\nonumber \\
	& &
	- \frac{1}{3!} v^{\phantom{\beta \gamma \delta} ; a}_{\beta \gamma \delta} \sigma^{; \beta} \sigma^{; \gamma} \sigma^{; \delta}
	+ \frac{4}{4!} v_{\beta \gamma \delta \epsilon} \sigma^{; \beta a} \sigma^{; \gamma} \sigma^{; \delta} \sigma^{; \epsilon}
	+ \frac{1}{4!} v^{\phantom{\beta \gamma \delta \epsilon} ; a}_{\beta \gamma \delta \epsilon} \sigma^{; \beta} \sigma^{; \gamma} \sigma^{; \delta} \sigma^{; \epsilon}
	\nonumber \\
	& &
	- \frac{5}{5!} v_{\beta \gamma \delta \epsilon \zeta} \sigma^{; \beta a} \sigma^{; \gamma} \sigma^{; \delta} \sigma^{; \epsilon} \sigma^{;\zeta}
	+ O\left( \sigma^{5/2} \right)
	\Big] d\tau '
\end{eqnarray}

Now, by definition, for a time-like geodesic, 
\begin{subequations}
\begin{eqnarray}
\label{eq:Synge}
\sigma &=& -\frac{1}{2}\left( \tau -\tau' \right)^2\\
\label{eq:SyngeDeriv}
\sigma ^{;a} &=& -\left( \tau -\tau' \right) u^{a}
\end{eqnarray}
\end{subequations}
where $u^{a}$ is the 4-velocity of the particle.

Also, the quantity $\sigma ^{;a b}$ can be written in terms of a covariant Taylor series expansion \cite{DeWitt:1960fc,Poisson:2003nc,Decanini:2005gt}:

\begin{equation}
\label{eq:Sigma_ab}
\sigma ^{;a b} = g^{a b}
	- \frac{1}{3}R^{a \phantom{\alpha} b}_{\phantom{a} \alpha \phantom{b} \beta}\sigma^{;\alpha} \sigma^{;\beta}
	+ \frac{1}{12}R^{a \phantom{\alpha} b}_{\phantom{a} \alpha \phantom{b} \beta ; \gamma} \sigma^{;\alpha} \sigma^{;\beta} \sigma^{;\gamma}
	- \left[ \frac{1}{60} R^{a \phantom{\alpha} b}_{\phantom{a} \alpha \phantom{b} \beta ; \gamma \delta} + \frac{1}{45}R^{a}_{\phantom{a} \alpha \rho \beta}R^{\rho \phantom{\gamma} b}_{\phantom{\rho} \gamma \phantom{b} \delta} \right] \sigma^{;\alpha} \sigma^{;\beta} \sigma^{;\gamma} \sigma^{;\delta}
	+ O\left( \left(\sigma^{;a}\right)^{5} \right)
\end{equation}

Substituting Eqs. (\ref{eq:Synge}), (\ref{eq:SyngeDeriv}) and (\ref{eq:Sigma_ab}) into Eq. (\ref{eq:SigmaExpandedIntegral}), the integrand becomes an expansion in powers of the proper time, $\tau - \tau'$:
\begin{eqnarray}
\label{eq:TauExpandedIntegral}
f^{a}_{\rm QL} &=& -q^2 \int_{\tau - \Delta \tau}^{\tau} \Big[
	A^{a} {\left( \tau - \tau ' \right)}^0 
	+ B^{a} {\left( \tau - \tau ' \right)}^1 
	+ C^{a} {\left( \tau - \tau ' \right)}^2 
	\nonumber \\
	& &
	~~~ ~~~
	+ D^{a} {\left( \tau - \tau ' \right)}^3 
	+ E^{a} {\left( \tau - \tau ' \right)}^4 
	+ O \left( \left( \tau - \tau ' \right)^5 \right)
	\Big] d\tau '
\end{eqnarray}
where
\begin{subequations}
\begin{eqnarray}
A^{a} &=& v^{;a} - v_{\beta} g^{\beta a} \label{eq:QLSFgenA} \\
B^{a} &=& v^{\phantom{\beta} ;a}_{\beta} u^{\beta} - v_{\beta \gamma} g^{\beta a} u^{\gamma} \label{eq:QLSFgenB} \\
C^{a} &=& \frac{1}{3} v_{\beta} R^{\beta \phantom{\gamma} a}_{\phantom{\beta} \gamma \phantom{a} \delta} u^{\gamma} u^{\delta}
					+ \frac{1}{2} v^{\phantom{\beta \gamma} ;a}_{\beta \gamma} u^{\beta} u^{\gamma}
					- \frac{1}{2} v_{\beta \gamma \delta} g^{\beta a} u^{\gamma} u^{\delta} \label{eq:QLSFgenC} \\
D^{a} &=& \frac{1}{12} v_{\beta} R^{\beta \phantom{\gamma} a}_{\phantom{\beta} \gamma \phantom{a} \delta ;\epsilon} u^{\gamma} u^{\delta} u^{\epsilon}
					+ \frac{1}{3} v_{\beta \gamma} R^{\beta \phantom{\delta} a}_{\phantom{\beta} \delta \phantom{a} \epsilon} u^{\gamma} u^{\delta} u^{\epsilon}
					+ \frac{1}{6} v^{\phantom{\beta \gamma \delta} ;a}_{\beta \gamma \delta} u^{\beta} u^{\gamma} u^{\delta}
					- \frac{1}{6} v_{\beta \gamma \delta \epsilon} g^{\beta a} u^{\gamma} u^{\delta} u^{\epsilon} \label{eq:QLSFgenD} \\
E^{a} &=& v_{\beta} \left( \frac{1}{60} R^{\beta \phantom{\gamma} a}_{\phantom{\beta} \gamma \phantom{a} \delta ;\epsilon \zeta} + \frac{1}{45} R^{\beta}_{\phantom{\beta} \gamma \rho \delta} R^{\rho \phantom{\epsilon} a}_{\phantom{\rho} \epsilon \phantom{a} \zeta} \right) u^{\gamma} u^{\delta} u^{\epsilon} u^{\zeta}
					+ \frac{1}{12} v_{\beta \gamma} R^{\beta \phantom{\delta} a}_{\phantom{\beta} \delta \phantom{a} \epsilon ;\zeta} u^{\gamma} u^{\delta} u^{\epsilon} u^{\zeta}
					\nonumber \\
					&&+ \frac{1}{6} v_{\beta \gamma \delta} R^{\beta \phantom{\epsilon} a}_{\phantom{\beta} \epsilon \phantom{a} \zeta} u^{\gamma} u^{\delta} u^{\epsilon} u^{\zeta}
					+ \frac{1}{24} v^{\phantom{\beta \gamma \delta \epsilon} ;a}_{\beta \gamma \delta \epsilon} u^{\beta} u^{\gamma} u^{\delta} u^{\epsilon}
					- \frac{1}{24} v_{\beta \gamma \delta \epsilon \zeta} g^{\beta a} u^{\gamma} u^{\delta} u^{\epsilon} u^{\zeta} \label{eq:QLSFgenE}
\end{eqnarray}
\end{subequations}
The coefficients of the powers of $\tau - \tau '$ are all local quantities at $x$, but the integral is in terms of the primed coordinates, $x'$, so the coefficients may be treated as constants while performing the integration. The result is that we only have to perform a trivial integral of powers of $\tau - \tau'$. Performing the integration gives us our final result, the quasi-local part of the self force in terms of an expansion in powers of the matching point, $\Delta \tau$:
\begin{equation}
\label{eq:QLSFgen}
f^{a}_{\rm QL} = -q^2 \left(
		   A^{a}{\Delta \tau}^1 
		+ \frac{1}{2} B^{a}{\Delta \tau}^2 
		+ \frac{1}{3} C^{a}{\Delta \tau}^3
		+ \frac{1}{4} D^{a}{\Delta \tau}^4
		+ \frac{1}{5} E^{a}{\Delta \tau}^5  + O \left( \Delta \tau^6 \right) \right)
\end{equation}

It is possible to simplify Eqs. (\ref{eq:QLSFgenA}) - (\ref{eq:QLSFgenE}) further by using the results of Appendix \ref{sec:cov-5} to rewrite all the odd order $v_{a_1 \dots a_p}$ terms in terms of the lower order even terms. It is then straightforward to substitute in for the $v_{a_1 \dots a_p}$ to get an expression for the quasi-local self-force in terms of only the scalar charge, field mass, coupling to the scalar background, 4-velocity, and products of Riemann tensor components of the background spacetime. However, we choose not to do so here as that would result in excessively lengthy expressions. Instead, we direct the reader to Ref. \cite{Decanini:2005gt} for a general expression for the $v_{n a_1 \dots a_p}$ for a scalar field and to Appendix \ref{sec:DeWittVacuum} where we give the expressions for the $v_{n a_1 \dots a_p}$ for massless scalar fields in vacuum spacetimes up to the orders required.

Once an expression has been obtained for the self-force using this method, it is straightforward to use Eqs. (\ref{eq:ma}) - (\ref{eq:dmdtau}) to obtain the equations of motion.

\section{Simplification of the quasi-local self-force for massless fields in vacuum space-times}
\label{sec:simplify}

Eq. (\ref{eq:QLSFgen}) is valid for any geodesic motion in any scalar field (massless or massive) in any background spacetime. At first it may appear that this result is quite difficult to work with due to the length of the expressions given in Ref. \cite{Decanini:2005gt} for the $v_{n a_1 \dots a_p}$. However, if we make the two assumptions that:
\begin{enumerate}
\item The scalar field is massless ($m_{\rm field}=0$)\label{item:massless}
\item The background spacetime is a vacuum spacetime ($R_{\alpha \beta}=0$)\label{item:vacuum}
\end{enumerate}
the majority of these terms vanish and we are left with the much more manageable expressions for the $v_{n a_1 \dots a_p}$ given in Appendix \ref{sec:DeWittVacuum}.  

In Appendix \ref{sec:DeWittVacuum}, we give the expansion coefficients 
$v_{n a_1 \dots a_p}$ up to $O \left( \sigma^{5/2} \right)$ for a massless field in a vacuum space-time.
We can see immediately from these expressions that all terms involving $v_{\alpha}$, $v_{\alpha \beta}$ and $v_{\alpha \beta \gamma}$ identically vanish. That leaves expressions involving only $v_{\alpha \beta \gamma \delta}$ and $v_{\alpha \beta \gamma \delta \epsilon}$. However, Eq. (\ref{eq:OddEven}) gives us a relation which allows us to express any odd order term (in this case, $v_{\alpha \beta \gamma \delta \zeta}$) in terms of all the lower order terms. This, along with the fact that $v_{a_1 \dots a_p}$ is totally symmetric, leads to a vastly simplified expression for the quasi-local self-force on a scalar particle following a geodesic in a vacuum background spacetime:

\begin{equation}
\label{eq:SimpForce}
f^{a}_{\rm QL} = q^2 \left(
		\frac{1}{4!} v_{\beta \gamma \delta \epsilon} g^{\beta a} u^\gamma u^\delta u^\epsilon {\Delta \tau}^4
		- \frac{1}{5!} \left( \frac{1}{2} v_{\gamma \delta \epsilon \zeta ;\beta} - 2 v_{\beta \gamma \delta \epsilon ;\zeta} \right)  g^{\beta a} u^\gamma u^\delta u^\epsilon u^\zeta {\Delta \tau}^5
		+ O \left( \Delta \tau^6 \right) \right)
\end{equation}

We see that a large number of terms in (\ref{eq:QLSFgen}) either identically vanish or cancel each other out. In fact, at the first three orders in $\Delta \tau$, we get no contribution to the quasi-local self-force whatsoever. Additionally, our fourth and fifth order coefficients take on a much simpler form than the general case given in Eqs. (\ref{eq:QLSFgenD}) and (\ref{eq:QLSFgenE}).

\section{Some Special Cases}
\label{sec:cases}
Expression (\ref{eq:SimpForce}) is a very general result, giving the quasi-local part of the self-force on a scalar charge following an arbitrary time-like geodesic in any 4-dimensional vacuum spacetime. While such a general result is of tremendous benefit in terms of flexibility, it is also interesting to examine some specific cases whose gravitational analog are of immediate physical interest. We will consider two well-known black hole spacetimes, Schwarzschild and Kerr, and some physically probable geodesics paths.

Although these results apply only to the non-physical scalar charge, we envisage that the techniques developed here could be applied to the more complicated gravitational case. This would allow us to work out the quasi-local part of the gravitational self-force. In conjunction with a mode-sum approximation of the far-field part of the integral, we would then be able to calculate the total self-force. Hence it would be possible to calculate the waveform corresponding to the gravitational radiation emitted from such a system. These waveforms are expected to be of great importance in the detection of gravitational wave sources by LISA \cite{LISA}, the space based gravitational wave observatory .

\subsection{Schwarzschild Space-time}
\label{sec:cases-schw}

We begin with Schwarzschild spacetime as that is the simpler of the two cases. The line-element of this spacetime is given by 
\begin{equation}
\label{eq:SchwMetric}
ds^2 = \left( 1-\frac{2M}{r} \right)^{-1} dr^2 + r^2 \left(d\theta ^2 + \sin ^2 \theta d\phi ^2 \right) - \left( 1-\frac{2M}{r} \right) dt^2
\end{equation}
From this, we calculate all necessary components of the Riemann tensor and its covariant derivatives and hence values for all the $v_{a_1 \dots a_p}$. While the calculations involved pose no conceptual difficulty, there is a great deal of scope for numerical slips. Fortunately, they are well suited to being done using a computer algebra system such as the GRTensor \cite{GRTensor} package for the Maple CAS \cite{Maple}.

We consider a particle with a general geodesic 4-velocity $u^\alpha$ travelling in this background spacetime. At first glance, it may appear that there are four independent components of this 4-velocity. However, because of the spherical symmetry of the problem, we may arbitrarily choose the orientation of the equatorial axis, $\theta$. By choosing this to line up with the 4-velocity (i.e. $\theta = \frac{\pi}{2}$), we can see immediately that $u^\theta=0$. Furthermore, the particle is following a time-like geodesic, so the normalization condition on the 4-velocity $u_{\alpha} u^\alpha = -1$ holds. This allows us to eliminate one of the three remaining components. In our case, we choose to eliminate $u^t$, which has the equation:

\begin{equation}
\label{eq:ut}
u^t = \sqrt{\left( \frac{r}{r-2M} \right) \left( 1 + \frac{r}{r-2M} \left( u^r \right)^2 + r^2 \left( u^\phi \right)^2 \right)}
\end{equation}

We are now left with just two independent components of the 4-velocity, $u^\phi$ and $u^r$. Substituting in to Eq. (\ref{eq:SimpForce}) the values for the known 4-velocity components $u^\theta$ and $u^t$ along with the Riemann tensor components, we get a general expression for the scalar self-force in terms of the scalar charge, $q$, the matching point, $\Delta \tau$, the mass of the Schwarzschild black hole, $M$, the radial distance of the scalar charge from the black hole, $r$ and the two independent components of the 4-velocity of the scalar charge, $u^r$ and $u^\phi$. We can then use Eqs. (\ref{eq:ma}) - (\ref{eq:dmdtau})  to obtain the quasi-local contribution to the equations of motion:

\begin{subequations}
\begin{eqnarray}
\label{eq:QLSF-SchwGen-r}
ma_{\rm QL}^{r} &=&
	\frac {3 q^2 M^2}{11200 r^{10}} \Bigg\{ 10 r^2 u^r \left[
	\left(\frac{r-2M}{r}\right) \left(1 + 11\left(ru^\phi\right)^2 + 13\left(ru^\phi\right)^4\right)
	+\left(u^r\right)^2 \left(1+8\left(ru^\phi\right)^2\right)
	\right]
	{\Delta \tau}^4 \nonumber \\
	& & 
	~~~~
	- \Bigg[
	\left(\frac{r-2M}{r}\right) \left(\left(20r-49M\right) + 8\left(11r-28M\right)\left(ru^\phi\right)^2 + 2\left(34r-89M\right)\left(ru^\phi\right)^4\right)
 	\nonumber \\
	& &~~~~~~~~
	+\left(u^r\right)^2 \left(\left(  4r-17M\right) - 8\left(31r-59M\right)\left(ru^\phi\right)^2-30\left(14r-29M\right)\left(ru^\phi\right)^4\right)
	\nonumber \\
	& &~~~~~~~~
	- 16r \left(u^r\right)^4	\left(1+15\left(ru^\phi\right)^2\right)
	\Bigg] {\Delta \tau}^5 + O \left( \Delta \tau ^6 \right) \Bigg\}\\
\label{eq:QLSF-SchwGen-theta}
ma_{\rm QL}^{\theta} &=& O \left( \Delta \tau ^6 \right)\\
\label{eq:QLSF-SchwGen-phi}
ma_{\rm QL}^{\phi} &=&
	\frac {3 q^2 M^2}{11200 r^{10}}  u^{\phi} \Bigg\{
	10r^2 \left[ 
	\left(\frac{r-2M}{r}\right) \left(7 + 20\left(ru^\phi\right)^2-13\left(ru^\phi\right)^4\right)
	+\left(u^r\right)^2 \left(5+8\left(ru^\phi\right)^2\right)
	\right]{\Delta \tau}^{4}
	\nonumber \\
	& &
	~~~~
	+ u^r \Bigg[
	3\left(68r-141M\right) + 48\left(13r-27M\right)\left(ru^\phi\right)^2 + 30\left(14r-29M\right)\left(ru^\phi\right)^4
 	\nonumber \\
	& &~~~~~~~~
	+ 48 r \left(u^r\right)^2\left(3+5\left(ru^\phi\right)^2\right)
	\Bigg] {\Delta \tau}^{5}
	+ O \left( \Delta \tau ^6 \right) \Bigg\}\\
\label{eq:QLSF-SchwGen-t}
ma_{\rm QL}^{t} &=&
	-\frac {3 q^2 M^2}{11200 r^{10}} \sqrt { \frac{r}{r-2M} \left( 1 + \left(ru^\phi\right)^2 + \frac{r}{r-2M} \left(u^r\right)^2\right)} \nonumber \\
	& & \times \Bigg\{
	10 r^2 \left[ 
	\left(\frac{r-2M}{r}\right) \left(7\left(ru^\phi\right)^2+13\left(ru^\phi\right)^4\right)
	+\left(u^r\right)^2 \left(1+8\left(ru^\phi\right)^2\right)
	\right] {\Delta \tau}^4
	\nonumber \\
	& &
	~~~~ + u^r \left[ 
	\left(20r-49M\right)  - 8\left(17r-31M\right)\left(ru^\phi\right)^2 - 30\left(14r-29M\right)\left(ru^\phi\right)^4
	- 16 r \left(u^r\right)^2\left(1+15\left(ru^\phi\right)^2\right)
	\right] {\Delta \tau}^5
	\nonumber \\
	& &
	~~~~ 
	+ O \left( \Delta \tau ^6 \right) \Bigg\}\\
\label{eq:dmdtau-SchwGen}
\frac{dm}{d\tau} &=&
	-\frac {3 q^2 M^2}{11200 r^{10}} \Bigg\{
	 5 r^2 \left[
	\left(\frac{r-2M}{r}\right) \left(5+28\left(ru^\phi\right)^2+26\left(ru^\phi\right)^4\right)
	+ 4 \left(u^r\right)^2 \left(1+4\left(ru^\phi\right)^2\right)
	\right] {\Delta \tau}^4
	\nonumber \\
	& &
	~~~~
	- 3 u^r \Bigg[ 
	\left(20r-41M\right) + 8\left(17r-35M\right)\left(ru^\phi\right)^2 +10\left(14r-29M\right)\left(ru^\phi\right)^4
 	+ 16 r \left(u^r\right)^2\left(1+ 5\left(ru^\phi\right)^2\right)
	\Bigg]  {\Delta \tau}^5
	\nonumber \\
	& &
	~~~~ 
	+ O \left( \Delta \tau ^6 \right) \Bigg\}
\end{eqnarray}
\end{subequations}
We see that the $\theta$-component of the 4-acceleration is $0$ as would be expected for motion in the equatorial plane. We also find that the other three components, $ma^r_{\rm QL}$, $ma^\phi_{\rm QL}$ and $ma^t_{\rm QL}$ all have their leading order terms at $O\left( \Delta \tau^4 \right)$ and also have terms at order $O\left( \Delta \tau^5 \right)$.

We may also express this result in an alternative way by making use of the constants of motion \cite{Hartle}:
\begin{subequations}
\begin{eqnarray}
e &=& \left( \frac{r - 2M}{r} \right) u^t\\
l &=& r^2 u^\phi\\
u^r &=& -\sqrt{e^2-1 - 2 V_{\rm{eff}}(r,e,l)}
\end{eqnarray}
\end{subequations}
where
\begin{equation}
 V_{\rm{eff}}(r,e,l) = -\frac{M}{r} + \frac{l^2}{2r^2} - \frac{Ml^2}{r^3}
\end{equation}
is an effective potential for the motion.

This gives a form for the equations of motion which will prove useful later (as a check on our result for Kerr spacetime):
\begin{subequations}
\begin{eqnarray}
\label{eq:ma-r-schw-el}
ma^r &=& {\frac {3 q^2 M^2}{11200{r}^{17}}}\left[
	\,{ -\sqrt{e^2-1 - 2 V_{\rm{eff}}}\,\left( 20\,{l}^{2}{r}^{7}-40\,{l}^{2}M{r}^{6}+50\,{l}^{4}{r}^{5}-100\,{l}^{4}M{r}^{4}+10\,{e}^{2}{r}^{9}+80\,{e}^{2}{l}^{2}{r}^{7} \right) {\Delta \tau}^{4}}\right.\nonumber \\
	& & ~~~
	+\,\left( -60\,{l}^{2}{r}^{6}+255\,{l}^{2}M{r}^{5}-270\,{l}^{2}{M}^{2}{r}^{4}-240\,{l}^{4}{r}^{4}+1008\,{l}^{4}M{r}^{3}-1056\,{l}^{4}{M}^{2}{r}^{2}-180\,{l}^{6}{r}^{2}+750\,{l}^{6}Mr \right. \nonumber \\
	& & ~~~~~~
	-780\,{l}^{6}{M}^{2}-36\,{e}^{2}{r}^{8}+81\,{e}^{2}M{r}^{7}-264\,{e}^{2}{l}^{2}{r}^{6}+552\,{e}^{2}{l}^{2}M{r}^{5}-60\,{e}^{2}{l}^{4}{r}^{4}+90\,{e}^{2}{l}^{4}M{r}^{3}+16\,{e}^{4}{r}^{8}\nonumber \\
	& & ~~~~~~
	\left. \left. +240\,{e}^{4}{l}^{2}{r}^{6} \right) {\Delta \tau}^{5} + O \left( \Delta \tau^6 \right)\right]\\
ma^\theta &=& O \left( \Delta \tau^6 \right)\\
ma^\phi &=& {\frac {3 q^2 M^2 l}{11200{r}^{16}}}\left[
	\left( 20\,{r}^{6}-40\,M{r}^{5}+70\,{l}^{2}{r}^{4}-140\,{l}^{2}M{r}^{3}+50\,{l}^{4}{r}^{2}-100\,{l}^{4}Mr+50\,{e}^{2}{r}^{6}+80\,{e}^{2}{l}^{2}{r}^{4} \right) {\Delta \tau}^{4} \right.\nonumber \\
	& & ~~~
	- \sqrt{e^2-1 - 2 V_{\rm{eff}}}\,\left( 60\,{r}^{5}-135\,M{r}^{4}+240\,{l}^{2}{r}^{3}-528\,{l}^{2}M{r}^{2}+180\,{l}^{4}r-390\,{l}^{4}M \right. \nonumber \\
	& & ~~~~~~
	\left. +144\,{e}^{2}{r}^{5}+240\,{e}^{2}{l}^{2}{r}^{3} \right) {\Delta \tau}^{5}
	\left. + O \left(\Delta \tau^6 \right)\right]\\
ma^t &=& {\frac {3 q^2 M^2 e}{11200\left( r-2\,M \right) {r}^{13}}} \times \nonumber \\
	& & ~~~
	\left[\left( -10\,{r}^{6}+20\,M{r}^{5}-20\,{l}^{2}{r}^{4}+40\,{l}^{2}M{r}^{3}+50\,{l}^{4}{r}^{2}-100\,{l}^{4}Mr+10\,{e}^{2}{r}^{6}+80\,{e}^{2}{l}^{2}{r}^{4} \right) {\Delta \tau}^{4} \right.\nonumber \\
	& & ~~~
	-\sqrt{e^2-1 - 2 V_{\rm{eff}}}\,\left( -36\,{r}^{5}+81\,M{r}^{4}-120\,{l}^{2}{r}^{3}+264\,{l}^{2}M{r}^{2}+180\,{l}^{4}r-390\,{l}^{4}M+16\,{e}^{2}{r}^{5} \right. \nonumber \\
	& & ~~~~~~
	+\left .240\,{e}^{2}{l}^{2}{r}^{3} \right) {\Delta \tau}^{5}
	\left. + O \left(\Delta \tau^6 \right)\right]\\
\label{eq:dmdtau-schw-el}
\frac{dm}{d\tau} &=&
	\frac {3 q^2 M^2}{11200{r}^{14}}  \big[
	\left( -5\,{r}^{6}+10\,M{r}^{5}-40\,{l}^{2}{r}^{4}+80\,{l}^{2}M{r}^{3}-50\,{l}^{4}{r}^{2}+100\,{l}^{4}Mr-20\,{e}^{2}{r}^{6}-80\,{e}^{2}{l}^{2}{r}^{4} \right) {\Delta \tau}^{4}\nonumber \\
	& &
	+ \sqrt{e^2-1 - 2 V_{\rm{eff}}} \left( 12\,{r}^{5}-27\,M{r}^{4}+120\,{l}^{2}{r}^{3}-264\,{l}^{2}M{r}^{2}+180\,{l}^{4}r-390\,{l}^{4}M+48\,{e}
^{2}{r}^{5}+240\,{e}^{2}{l}^{2}{r}^{3} \right) {\Delta \tau}^{5}\nonumber \\
	& &
	+ O \left( \Delta \tau^6\right) \big]
\end{eqnarray}
\end{subequations}

These expressions give us the quasi-local contribution to the equations of motion for a particle following any geodesic path in a Schwarzschild background. It is also interesting to look at some specific particle paths. In particular, in order to check our results, we consider three cases for which results using other methods are already available in the literature \cite{Anderson:2005gb,Anderson:2005ds}\footnote{Ref. \cite{Anderson:2005ds} also considers the case of a static particle, but in that case the motion is non-geodesic, so we leave the analysis for later work.}.
\begin{enumerate}
\item A particle following a circular geodesic
\item A particle under radial in-fall
\item A particle under radial in-fall from rest
\end{enumerate}

\subsubsection{Circular geodesic in Schwarzschild}
For a particle following a circular geodesic in Schwarzschild spacetime, we have the condition $u^r = 0$ along with the previous condition $u^\theta = 0$. Furthermore, for a circular geodesic orbit $u^\phi$ is uniquely determined by the radius, $r$, angle, $\theta$ and mass, $M$ of the Schwarzschild black hole:
\begin{equation}
\label{eq:SchwCirc-u-phi}
u^\phi = \frac{1}{r} \sqrt{\frac{M}{r-3M}}
\end{equation}
Substituting (\ref{eq:SchwCirc-u-phi}) into (\ref{eq:ut}), we also obtain the expression for $u^t$ in terms of the radius, $r$, and mass, $M$ of the Schwarzschild black hole:
\begin{equation}
\label{eq:SchwCirc-u-r}
u^t = \sqrt{\frac{r}{r-3M}}
\end{equation}

We can now use these values for the 4-velocity in Eqs. (\ref{eq:QLSF-SchwGen-r}) - (\ref{eq:dmdtau-SchwGen}). In doing so, the expressions simplify significantly and we obtain a result for the  quasi-local contribution to the equations of motion for a scalar particle following a circular geodesic in terms of $q$, $r$ and $M$ alone:
\begin{subequations}
\begin{eqnarray}
ma_{\rm QL}^{r} &=& - \frac{3 q^2 M^2 \left( r-2M \right) \left( 20 r^3 - 81 M r^2 + 54 r M^2 + 53 M^3 \right)}{11200 \left( r-3M \right)^2 r^{11}} \Delta \tau^5 + O \left( \Delta \tau ^6 \right) \\
ma_{\rm QL}^{\theta} &=& O \left( \Delta \tau ^6 \right) \\
ma_{\rm QL}^{\phi} &=& \frac{3 q^2 M^2 \left( r-2M \right)^2 \left( 7r-8M \right)}{1120 \left( r-3M \right)^2r^{10} } \sqrt{\frac{M}{r-3M}} \Delta \tau^4 + O \left( \Delta \tau ^6 \right) \\
ma_{\rm QL}^{t} &=& \frac{3 q^2 M^3 \left( r-2M \right) \left( 7r-8M \right)}{1120 \left( r-3M \right)^2 r^{9}} \sqrt{\frac{r}{r-3M}} \Delta \tau^4 + O \left( \Delta \tau ^6 \right) \\
\frac{dm}{d\tau} &=& \frac { 3 q^2 M^2 \left( r-2M \right) \left( 13M^2 + 2Mr-  5r^2 \right) }{2240 \left( r-3M \right) ^2 r^9} \Delta \tau^{4} + O \left( \Delta \tau ^6 \right) 
\end{eqnarray}
\end{subequations}

Our results exactly match Eqs. (\ref{eq:rCoordCirc}) - (\ref{eq:dmdtauCoordCirc}) containing our corrected version of the results of Anderson and Wiseman\cite{Anderson:2005gb}.

\subsubsection{Radial geodesic in Schwarzschild}
A particle following a radial geodesic in Schwarzschild spacetime has the conditions on the 4-velocity that $u^\phi = 0$ in addition to the previous condition $u^\theta = 0$. We can therefore write Eq. (\ref{eq:ut}) as:
\begin{equation}
\label{eq:SchwRad-u-t-implicit}
u^t = \sqrt{\frac{r}{r-2M}\left( 1+ \frac{r}{r-2M} \left({u^r}\right)^2 \right) }
\end{equation}
This leaves us with one independent component of the 4-velocity, $u^r$. 
Substituting for $u^\phi$ and $u^t$ into (\ref{eq:QLSF-SchwGen-r}) - (\ref{eq:dmdtau-SchwGen}), the expression simplifies significantly and we obtain a result which is only slightly more complicated than for the case of a particle following a circular geodesic. Our expression for the  quasi-local contribution to the equations of motion is now in terms of the scalar charge, $q$ radius, $r$, the mass, $M$ and the radial 4-velocity of the particle, $u^r$:
\begin{subequations}
\begin{eqnarray}
\label{eq:RadialQLSF-r}
ma_{\rm QL}^{r} &=&
	\frac{3 q^2 M^2 }{11200 r^{11}} \left(r-2M+r \left(u^r\right)^2 \right) \left[ 10 r^2 u^r \Delta \tau^4 
	+ \left(49M - 20r + 16r\left(u^r\right)^{2}\right)  \Delta \tau^5
	+ O \left( \Delta \tau ^6 \right) \right] \\
ma_{\rm QL}^{\theta} &=& O \left( \Delta \tau ^6 \right) \\
ma_{\rm QL}^{\phi} &=& O \left( \Delta \tau ^6 \right) \\
ma_{\rm QL}^{t} &=&
	\frac{-3 q^2 M^2 u^r }{11200 \left(r-2M\right) r^{9}} \sqrt{\frac{r-2M}{r}+\left({u^r}\right)^2} \left[
	10 r^2 u^r \Delta \tau^4 + \left( 20r - 49M - 16r \left( u^r \right)^2 \right) \Delta \tau^5
	+ O \left( \Delta \tau ^6 \right)  \right] \\
\label{eq:Radialdmdtau}
\frac{dm}{d\tau} &=&
	\frac{-3 q^2 M^2}{11200 r^{10}} \left[ 5r \left(5r - 10M + 4 r \left(u^r\right)^2 \right) \Delta \tau^4 
	+u^r \left(60r - 123M + 48r \left(u^r\right)^2\right)\Delta \tau^5
	+ O \left( \Delta \tau ^6 \right) \right]
\end{eqnarray}
\end{subequations}

From angular momentum considerations, we would expect a radially moving particle in a spherically symmetric background to experience no acceleration in the $\theta$ or $\phi$ directions. This is confirmed by our results.

\subsubsection{Radial geodesic in Schwarzschild: Starting from rest}
\label{sec:rad-rest}
In order to calculate the quasi-local contribution to the equations of motion for a particle under radial infall from rest, we begin with expressions (\ref{eq:RadialQLSF-r}) - (\ref{eq:Radialdmdtau}) for a particle under general radial infall. For the sake of simplicity, we will take $\tau=0$ to be the proper time at which the particle is released from rest at a point $r=r_0$. We also assume that at proper time $\tau = \Delta \tau$, the particle will be at the radial point $r$, travelling radially inwards with 4-velocity $u^r$. In this case, clearly expressions (\ref{eq:RadialQLSF-r}) - (\ref{eq:Radialdmdtau}) hold. However, we must now obtain an expression for $u^r = u^r \left(\Delta\tau\right)$ satisfying the initial condition $u^r \left(0\right) = 0$.

We begin with an expression for the radial 4-velocity in radial geodesic motion starting from rest at $r_0$ \cite{Chandrasekhar}:
\begin{equation}
\label{eq:ur-rad-rest}
\left(u^r\right)^2 = \left(\frac{dr}{d\tau}\right)^2 = 2M \left( \frac{1}{r} - \frac{1}{r_0} \right)
\end{equation}

Assuming that the particle has not travelled too far (i.e. $(r-r_0)/r_0 \ll 1$), this equation may be approximately integrated with respect to $r$ from $r_0$ up to $r$, giving an approximate expression for $r_0$:

\begin{equation}
r_0 = r + \frac{M \Delta \tau^2}{2 r_0^2} + O \left(\Delta\tau^3\right) = r + \frac{M \Delta \tau^2}{2 r^2} + O \left(\Delta\tau^3\right) 
\end{equation}

Using this in (\ref{eq:ur-rad-rest}), we get an approximate expression for $u^r$ (we have taken the negative square root since we are considering infalling particles):
\begin{equation}
u^r = -\frac{M\Delta\tau}{r^2} + O \left( \Delta \tau ^2 \right) 
\end{equation}

We can then use this expression in (\ref{eq:RadialQLSF-r}) - (\ref{eq:Radialdmdtau}) to give the quasi-local contribution to the equations of motion for a particle falling radially inward from rest in Schwarzschild spacetime: 
\begin{subequations}
\begin{eqnarray}
\label{eq:RadialRestQLSF}
ma_{\rm QL}^{r} &=&
	- \frac {3 q^2 M^2}{11200 r^{10}}   \left( \frac{r-2M}{r} \right) \left( 20r-39M \right) {\Delta \tau}^5 + O \left( \Delta \tau ^6 \right) \\
ma_{\rm QL}^{\theta} &=& O \left( \Delta \tau ^6 \right) \\
ma_{\rm QL}^{\phi} &=& O \left( \Delta \tau ^6 \right) \\
ma_{\rm QL}^{t} &=& O \left( \Delta \tau ^6 \right) \\
\label{eq:RadialRestdmdtau}\frac{dm}{d\tau} &=&
	- \frac {3 q^2 M^2}{448 r^8} \left(\frac{r - 2M}{r} \right) {\Delta \tau}^4 + O \left( \Delta \tau ^6 \right) 
\end{eqnarray}
\end{subequations}

As explained in Section \ref{sec:coord-radial}, this result is in agreement with those of Ref. \cite{Anderson:2005ds}.

\subsection{Kerr Space-time}
\label{sec:cases-kerr}
Next, we look at the much more complicated but physically more interesting Kerr spacetime. In Boyer-Lindquist coordinates, the line-element is
\begin{equation}
\label{eq:KerrMetric}
ds^2 = - \left(1-\frac{2Mr}{\Sigma}\right)dt^2 
			- \frac{4aMr\sin^2\theta}{\Sigma}dtd\phi
			+ \frac{\Sigma}{\Delta}dr^2
			+ \Sigma d\theta^2
			+ \left(\Delta+\frac{2Mr(r^2+a^2)} {\Sigma}\right) \sin^2\theta d\phi^2
\end{equation}
where
\begin{eqnarray}
\Sigma &=& r^2+a^2\cos^2\theta\\
\Delta &=& r^2-2Mr+a^2
\end{eqnarray}
Although it is possible to use a computer algebra package to obtain a totally general solution for the self-force in Kerr spacetime, the length of the resulting expressions are too unwieldy to be of much benefit in printed form. For this reason, we will only focus here on equatorial motion, so that $\theta = \frac{\pi}{2}$, keeping in mind that more general results may be obtained using the same methodology.

As a result of the axial symmetry of the metric, motion in the equatorial plane in Kerr spacetime will always remain in the equatorial plane, i.e. $u^\theta = 0$. Additionally, the motion is parametrized by two quantities analogous to the Schwarzschild case: the conserved energy per unit mass, $e$ and the angular momentum per unit mass along the symmetry axis, $l$. We can therefore write the three non-zero components of the 4-velocity in terms of these conserved quantities \cite{Hartle}:

\begin{subequations}
\begin{eqnarray}
\label{eq:kerr-ur}
u^r = \frac{dr}{d\tau} &=& - \sqrt{e^2-1 - 2 V_{\rm{eff}}(r,e,l)}\\
\label{eq:kerr-uphi}
u^\phi = \frac{d\phi}{d\tau} &=& \frac{1}{\Delta}\left[\left(\frac{r-2M}{r}\right)l + \frac{2Ma}{r}e\right]\\
\label{eq:kerr-ut}
u^t = \frac{dt}{d\tau} &=& \frac{1}{\Delta} \left[\left(r^2 + a^2 + \frac{2Ma^2}{r}\right)e - \frac{2Ma}{r}l\right] 
\end{eqnarray}
\end{subequations}
where
\begin{equation}
V_{\rm{eff}}(r,e,l) = -\frac{M}{r} + \frac{l^2-a^2\left(e^2-1\right)}{2r^2}-\frac{M\left(l-ae\right)^2}{r^3}
\end{equation}
is the effective potential for equatorial motion.

Substituting these values into Eq. (\ref{eq:SimpForce}) and computing the relevant components of the Hadamard/DeWitt coefficients, we arrive at a lengthy but general set of expressions for the equations of motion for a particle following a geodesic in the equatorial plane in Kerr spacetime:
\begin{subequations}
\begin{eqnarray}
ma^r &=& 
	-{\frac {3 q^2 M^2}{11200 r^{14}}} \sqrt{e^2-1 - 2 V_{\rm{eff}}} \Bigg[
	\left( 20\,{l}^{2}{r}^{4}-40\,{l}^{2}M{r}^{3}+50\,{l}^{4}{r}^{2}-100\,{l}^{4}M{r}^{1}+10\,{e}^{2}{r}^{6}+80\,{e}^{2}{l}^{2}{r}^{4} \right) \nonumber \\
	& & ~~~
	+ a\left( -120\,el{r}^{4}+80\,elM{r}^{3}-600\,e{l}^{3}{r}^{2}+400\,e{l}^{3}M{r}^{1}-160\,{e}^{3}l{r}^{4} \right) \nonumber \\
	& & ~~~
	+ a^2 \left( 150\,{l}^{2}{r}^{2}+500\,{l}^{4}+100\,{e}^{2}{r}^{4}-40\,{e}^{2}M{r}^{3}+1500\,{e}^{2}{l}^{2}{r}^{2}-600\,{e}^{2}{l}^{2}M{r}^{1}+80\,{e}^{4}{r}^{4} \right)\nonumber \\
	& & ~~~
	+ a^3\left( -300\,el{r}^{2}-2000\,e{l}^{3}-1400\,{e}^{3}l{r}^{2}+400\,{e}^{3}lM{r}^{1} \right)\nonumber \\
	& & ~~~
	+ a^4 \left( 150\,{e}^{2}{r}^{2}+3000\,{e}^{2}{l}^{2}+450\,{e}^{4}{r}^{2}-100\,{e}^{4}M{r}^{1} \right) \nonumber \\
	& & ~~~
	- a^5 \left( 2000 \,{e}^{3}l \right)
	+ a^6 \left( 500 \,{e}^{4} \right) \Bigg]{\Delta \tau}^{4} \nonumber \\
	& & 
	-{\frac {3 q^2 M^2}{11200 r^{18}}}\Bigg[
	\big( 60\,{l}^{2}{r}^{7}-255\,{l}^{2}M{r}^{6}+270\,{l}^{2}{M}^{2}{r}^{5}+240\,{l}^{4}{r}^{5}-1008\,{l}^{4}M{r}^{4}+1056\,{l}^{4}{M}^{2}{r}^{3}+180\,{l}^{6}{r}^{3} \nonumber \\
	& & ~~~~~~~~
	-750\,{l}^{6}M{r}^{2}+780\,{l}^{6}{M}^{2}r+36\,{e}^{2}{r}^{9}-81\,{e}^{2}M{r}^{8}+264\,{e}^{2}{l}^{2}{r}^{7}-552\,{e}^{2}{l}^{2}M{r}^{6}+60\,{e}^{2}{l}^{4}{r}^{5} \nonumber \\
	& & ~~~~~~~~
	-90\,{e}^{2}{l}^{4}M{r}^{4}-16\,{e}^{4}{r}^{9}-240\,{e}^{4}{l}^{2}{r}^{7} \big) \nonumber \\
	& & ~~~
	+ a \big( -432\,el{r}^{7}+1176\,elM{r}^{6}-540\,el{M}^{2}{r}^{5}-2760\,e{l}^{3}{r}^{5}+7728\,e{l}^{3}M{r}^{4}-4224\,e{l}^{3}{M}^{2}{r}^{3}-2160\,e{l}^{5}{r}^{3} \nonumber \\
	& & ~~~~~~~~
	+6600\,e{l}^{5}M{r}^{2}-4680\,e{l}^{5}{M}^{2}r-368\,{e}^{3}l{r}^{7}+1104\,{e}^{3}lM{r}^{6}+1680\,{e}^{3}{l}^{3}{r}^{5}+360\,{e}^{3}{l}^{3}M{r}^{4}+480\,{e}^{5}l{r}^{7} \big) \nonumber \\
	& & ~~~
	+ a^2 \big( 660\,{l}^{2}{r}^{5}-1368\,{l}^{2}M{r}^{4}+3000\,{l}^{4}{r}^{3}-6120\,{l}^{4}M{r}^{2}+2100\,{l}^{6}r-4200\,{l}^{6}M+456\,{e}^{2}{r}^{7} \nonumber \\
	& & ~~~~~~~~
	-921\,{e}^{2}M{r}^{6}+270\,{e}^{2}{M}^{2}{r}^{5}+6720\,{e}^{2}{l}^{2}{r}^{5}-17136\,{e}^{2}{l}^{2}M{r}^{4}+6336\,{e}^{2}{l}^{2}{M}^{2}{r}^{3}+3120\,{e}^{2}{l}^{4}{r}^{3} \nonumber \\
	& & ~~~~~~~~
	-21750\,{e}^{2}{l}^{4}M{r}^{2}+11700\,{e}^{2}{l}^{4}{M}^{2}r+104\,{e}^{4}{r}^{7}-552\,{e}^{4}M{r}^{6}-5400\,{e}^{4}{l}^{2}{r}^{5}-540\,{e}^{4}{l}^{2}M{r}^{4}-240\,{e}^{6}{r}^{7} \big) \nonumber \\
	& & ~~~
	+ a^3 \big( -1800\,el{r}^{5}+2736\,elM{r}^{4}-13600\,e{l}^{3}{r}^{3}+24480\,e{l}^{3}M{r}^{2}-8400\,e{l}^{5}r+25200\,e{l}^{5}M-6120\,{e}^{3}l{r}^{5} \nonumber \\
	& & ~~~~~~~~
	+15120\,{e}^{3}lM{r}^{4}-4224\,{e}^{3}l{M}^{2}{r}^{3}+5520\,{e}^{3}{l}^{3}{r}^{3}+36000\,{e}^{3}{l}^{3}M{r}^{2}-15600\,{e}^{3}{l}^{3}{M}^{2}r+5520\,{e}^{5}l{r}^{5} \nonumber \\
	& & ~~~~~~~~
	+360\,{e}^{5}lM{r}^{4} \big) \nonumber \\
	& & ~~~
	+ a^4 \big( 720\,{l}^{2}{r}^{3}+2800\,{l}^{4}r+1140\,{e}^{2}{r}^{5}-1368\,{e}^{2}M{r}^{4}+22800\,{e}^{2}{l}^{2}{r}^{3}-36720\,{e}^{2}{l}^{2}M{r}^{2}+10500\,{e}^{2}{l}^{4}r \nonumber \\
	& & ~~~~~~~~
	-63000\,{e}^{2}{l}^{4}M+1920\,{e}^{4}{r}^{5}-4704\,{e}^{4}M{r}^{4}+1056\,{e}^{4}{M}^{2}{r}^{3}-16380\,{e}^{4}{l}^{2}{r}^{3}-32250\,{e}^{4}{l}^{2}M{r}^{2} \nonumber \\
	& & ~~~~~~~~
	+11700\,{e}^{4}{l}^{2}{M}^{2}r-1860\,{e}^{6}{r}^{5}-90\,{e}^{6}M{r}^{4} \big)  \nonumber \\
	& & ~~~
	+ a^5 \big( -1440\,el{r}^{3}-11200\,e{l}^{3}r-16800\,{e}^{3}l{r}^{3}+24480\,{e}^{3}lM{r}^{2}+84000\,{e}^{3}{l}^{3}M+13440\,{e}^{5}l{r}^{3} \nonumber \\
	& & ~~~~~~~~
	+15000\,{e}^{5}lM{r}^{2}-4680\,{e}^{5}l{M}^{2}r \big) \nonumber \\
	& & ~~~
	+ a^6 \big( 720\,{e}^{2}{r}^{3}+16800\,{e}^{2}{l}^{2}r+4600\,{e}^{4}{r}^{3}-6120\,{e}^{4}M{r}^{2}-10500\,{e}^{4}{l}^{2}r-63000\,{e}^{4}{l}^{2}M-3720\,{e}^{6}{r}^{3} \nonumber \\
	& & ~~~~~~~~
	-2850\,{e}^{6}M{r}^{2}+780\,{e}^{6}{M}^{2}r \big) \nonumber \\
	& & ~~~
	+ a^7 \left( -11200\,{e}^{3}lr+8400\,{e}^{5}lr+25200\,{e}^{5}lM \right)\nonumber \\
	& & ~~~
	+ a^8 \left( 2800\,{e}^{4}r-2100\,{e}^{6}r-4200\,{e}^{6}M \right)
	\Bigg] {\Delta \tau}^{5} + O \left( \Delta \tau ^6 \right)\\
ma^\theta &=& O\left( \Delta \tau^6 \right)\\
ma^\phi &=&
	-\frac{3 q^2 M^2}{11200 \left( {r}^{2}-2\,Mr+{a}^{2} \right) {r}^{16}} \Bigg[
	\big( -20\,l{r}^{8}+80\,lM{r}^{7}-80\,l{M}^{2}{r}^{6}-70\,{l}^{3}{r}^{6}+280\,{l}^{3}M{r}^{5}-280\,{l}^{3}{M}^{2}{r}^{4}\nonumber \\
	& & ~~~~~~~~
	-50\,{l}^{5}{r}^{4}+200\,{l}^{5}M{r}^{3}-200\,{l}^{5}{M}^{2}{r}^{2}-50\,{e}^{2}l{r}^{8}+100\,{e}^{2}lM{r}^{7}-80\,{e}^{2}{l}^{3}{r}^{6}+160\,{e}^{2}{l}^{3}M{r}^{5} \big)\nonumber \\
	& & ~~~
	+ a \big( 60\,e{r}^{8}-160\,eM{r}^{7}+80\,e{M}^{2}{r}^{6}+570\,e{l}^{2}{r}^{6}-1560\,e{l}^{2}M{r}^{5}+840\,e{l}^{2}{M}^{2}{r}^{4}+600\,e{l}^{4}{r}^{4}\nonumber \\
	& & ~~~~~~~~
	-1700\,e{l}^{4}M{r}^{3}+1000\,e{l}^{4}{M}^{2}{r}^{2}+40\,{e}^{3}{r}^{8}-100\,{e}^{3}M{r}^{7}+160\,{e}^{3}{l}^{2}{r}^{6}-480\,{e}^{3}{l}^{2}M{r}^{5} \big) \nonumber \\
	& & ~~~
	+ a^2 \big( -170\,l{r}^{6}+340\,lM{r}^{5}-700\,{l}^{3}{r}^{4}+1400\,{l}^{3}M{r}^{3}-500\,{l}^{5}{r}^{2}+1000\,{l}^{5}Mr-890\,{e}^{2}l{r}^{6}+2280\,{e}^{2}lM{r}^{5}\nonumber \\
	& & ~~~~~~~~
	-840\,{e}^{2}l{M}^{2}{r}^{4}-1500\,{e}^{2}{l}^{3}{r}^{4}+4800\,{e}^{2}{l}^{3}M{r}^{3}-2000\,{e}^{2}{l}^{3}{M}^{2}{r}^{2}-80\,{e}^{4}l{r}^{6}+480\,{e}^{4}lM{r}^{5} \big) \nonumber \\
	& & ~~~
	+ a^3 \big( 210\,e{r}^{6}-340\,eM{r}^{5}+2250\,e{l}^{2}{r}^{4}-4200\,e{l}^{2}M{r}^{3}+2000\,e{l}^{4}{r}^{2}-5000\,e{l}^{4}Mr+390\,{e}^{3}{r}^{6}\nonumber \\
	& & ~~~~~~~~
	-1000\,{e}^{3}M{r}^{5}+280\,{e}^{3}{M}^{2}{r}^{4}+1400\,{e}^{3}{l}^{2}{r}^{4}-6200\,{e}^{3}{l}^{2}M{r}^{3}+2000\,{e}^{3}{l}^{2}{M}^{2}{r}^{2}-160\,{e}^{5}M{r}^{5} \big)\nonumber \\
	& & ~~~
	+ a^4\big( -150\,l{r}^{4}-500\,{l}^{3}{r}^{2}-2400\,{e}^{2}l{r}^{4}+4200\,{e}^{2}lM{r}^{3}-3000\,{e}^{2}{l}^{3}{r}^{2}+10000\,{e}^{2}{l}^{3}Mr-450\,{e}^{4}l{r}^{4}\nonumber \\
	& & ~~~~~~~~
	+3800\,{e}^{4}lM{r}^{3}-1000\,{e}^{4}l{M}^{2}{r}^{2} \big) \nonumber \\
	& & ~~~
	+ a^5 \big( 150\,e{r}^{4}+1500\,e{l}^{2}{r}^{2}+850\,{e}^{3}{r}^{4}-1400\,{e}^{3}M{r}^{3}+2000\,{e}^{3}{l}^{2}{r}^{2}-10000\,{e}^{3}{l}^{2}Mr-900\,{e}^{5}M{r}^{3}\nonumber \\
	& & ~~~~~~~~
	+200\,{e}^{5}{M}^{2}{r}^{2} \big) \nonumber \\
	& & ~~~
	+ a^6 \big( -1500\,{e}^{2}l{r}^{2}-500\,{e}^{4}l{r}^{2}+5000\,{e}^{4}lMr \big) \nonumber \\
	& & ~~~
	+ a^7 \big( 500\,{e}^{3}{r}^{2}-1000\,{e}^{5}Mr \big)
	\Bigg] {\Delta \tau}^{4}  \nonumber \\
	& & 
	-\frac {3 q^2 M^2}{11200\left( {r}^{2}-2\,Mr+{a}^{2} \right) {r}^{16}} \sqrt{e^2-1 - 2 V_{\rm{eff}}} \Bigg[
	\big( 60\,l{r}^{7}-255\,lM{r}^{6}+270\,l{M}^{2}{r}^{5}+240\,{l}^{3}{r}^{5}\nonumber \\
	& & ~~~~~~~~
	-1008\,{l}^{3}M{r}^{4}+1056\,{l}^{3}{M}^{2}{r}^{3}	+180\,{l}^{5}{r}^{3}-750\,{l}^{5}M{r}^{2}+780\,{l}^{5}{M}^{2}r+144\,{e}^{2}l{r}^{7}-288\,{e}^{2}lM{r}^{6}\nonumber \\
	& & ~~~~~~~~
	+240\,{e}^{2}{l}^{3}{r}^{5}-480\,{e}^{2}{l}^{3}M{r}^{4} \big) \nonumber \\
	& & ~~~ 
	+a \big( -192\,e{r}^{7}+528\,eM{r}^{6}-270\,e{M}^{2}{r}^{5}-2040\,e{l}^{2}{r}^{5}+5688\,e{l}^{2}M{r}^{4}-3168\,e{l}^{2}{M}^{2}{r}^{3}-2160\,e{l}^{4}{r}^{3} \nonumber \\
	& & ~~~~~~~~
	+6240\,e{l}^{4}M{r}^{2}-3900\,e{l}^{4}{M}^{2}r-128\,{e}^{3}{r}^{7}+288\,{e}^{3}M{r}^{6}-480\,{e}^{3}{l}^{2}{r}^{5}+1440\,{e}^{3}{l}^{2}M{r}^{4} \big) \nonumber \\
	& & ~~~
	+ a^2 \big( 660\,l{r}^{5}-1368\,lM{r}^{4}+3000\,{l}^{3}{r}^{3}-6120\,{l}^{3}M{r}^{2}+2100\,{l}^{5}r-4200\,{l}^{5}M+3360\,{e}^{2}l{r}^{5}-8352\,{e}^{2}lM{r}^{4} \nonumber \\
	& & ~~~~~~~~
	+3168\,{e}^{2}l{M}^{2}{r}^{3}+5400\,{e}^{2}{l}^{3}{r}^{3}-17460\,{e}^{2}{l}^{3}M{r}^{2}+7800\,{e}^{2}{l}^{3}{M}^{2}r+240\,{e}^{4}l{r}^{5}-1440\,{e}^{4}lM{r}^{4} \big) \nonumber \\
	& & ~~~
	+ a^3 \big( -840\,e{r}^{5}+1368\,eM{r}^{4}-10080\,e{l}^{2}{r}^{3}+18360\,e{l}^{2}M{r}^{2}-8400\,e{l}^{4}r+21000\,e{l}^{4}M-1560\,{e}^{3}{r}^{5} \nonumber \\
	& & ~~~~~~~~
	+3672\,{e}^{3}M{r}^{4}-1056\,{e}^{3}{M}^{2}{r}^{3}-5040\,{e}^{3}{l}^{2}{r}^{3}+22440\,{e}^{3}{l}^{2}M{r}^{2}-7800\,{e}^{3}{l}^{2}{M}^{2}r+480\,{e}^{5}M{r}^{4} \big) \nonumber \\
	& & ~~~
	+ a^4 \big( 720\,l{r}^{3}+2800\,{l}^{3}r+11160\,{e}^{2}l{r}^{3}-18360\,{e}^{2}lM{r}^{2}+12600\,{e}^{2}{l}^{3}r-42000\,{e}^{2}{l}^{3}M+1620\,{e}^{4}l{r}^{3} \nonumber \\
	& & ~~~~~~~~
	-13710\,{e}^{4}lM{r}^{2}+3900\,{e}^{4}l{M}^{2}r \big)\nonumber \\
	& & ~~~
	+ a^5 \big( -720\,e{r}^{3}-8400\,e{l}^{2}r-4080\,{e}^{3}{r}^{3}+6120\,{e}^{3}M{r}^{2}-8400\,{e}^{3}{l}^{2}r+42000\,{e}^{3}{l}^{2}M+3240\,{e}^{5}M{r}^{2} \nonumber \\
	& & ~~~~~~~~
	-780\,{e}^{5}{M}^{2}r \big)\nonumber \\
	& & ~~~
	+ a^6 \big( 8400\,{e}^{2}lr+2100\,{e}^{4}lr-21000\,{e}^{4}lM \big)\nonumber \\
	& & ~~~
	+ a^7 \big( -2800\,{e}^{3}r+4200\,{e}^{5}M \big) \Bigg] {\Delta \tau}^{5} + O \left( \Delta \tau ^6 \right)\\
ma^t &=&
	-{\frac {3 q^2 M^2}{11200 \left( {r}^{2}-2\,Mr+{a}^{2} \right) {r}^{16}}} \Bigg[
	\big( 10\,e{r}^{10}-20\,eM{r}^{9}+20\,e{l}^{2}{r}^{8}-40\,e{l}^{2}M{r}^{7}-50\,e{l}^{4}{r}^{6}+100\,e{l}^{4}M{r}^{5}\nonumber \\
	& & ~~~~~~~~
	-10\,{e}^{3}{r}^{10}-80\,{e}^{3}{l}^{2}{r}^{8} \big) \nonumber \\
	& & ~~~
	+ a \big( -60\,l{r}^{8}+160\,lM{r}^{7}-80\,l{M}^{2}{r}^{6}-150\,{l}^{3}{r}^{6}+440\,{l}^{3}M{r}^{5}-280\,{l}^{3}{M}^{2}{r}^{4}+100\,{l}^{5}M{r}^{3}\nonumber \\
	& & ~~~~~~~~
	-200\,{l}^{5}{M}^{2}{r}^{2}+180\,{e}^{2}lM{r}^{7}+600\,{e}^{2}{l}^{3}{r}^{6}-240\,{e}^{2}{l}^{3}M{r}^{5}+160\,{e}^{4}l{r}^{8} \big) \nonumber \\
	& & ~~~
	+ a^2 \big( 110\,e{r}^{8}-240\,eM{r}^{7}+80\,e{M}^{2}{r}^{6}+620\,e{l}^{2}{r}^{6}-2040\,e{l}^{2}M{r}^{5}+840\,e{l}^{2}{M}^{2}{r}^{4}-550\,e{l}^{4}{r}^{4}\nonumber \\
	& & ~~~~~~~~
	-1200\,e{l}^{4}M{r}^{3}+1000\,e{l}^{4}{M}^{2}{r}^{2}-30\,{e}^{3}{r}^{8}-140\,{e}^{3}M{r}^{7}-1580\,{e}^{3}{l}^{2}{r}^{6}+120\,{e}^{3}{l}^{2}M{r}^{5}-80\,{e}^{5}{r}^{8} \big)\nonumber \\
	& & ~~~
	+ a^3 \big( -210\,l{r}^{6}+340\,lM{r}^{5}-650\,{l}^{3}{r}^{4}+1400\,{l}^{3}M{r}^{3}+1000\,{l}^{5}Mr-750\,{e}^{2}l{r}^{6}+2760\,{e}^{2}lM{r}^{5}\nonumber \\
	& & ~~~~~~~~
	-840\,{e}^{2}l{M}^{2}{r}^{4}+2600\,{e}^{2}{l}^{3}{r}^{4}+3800\,{e}^{2}{l}^{3}M{r}^{3}-2000\,{e}^{2}{l}^{3}{M}^{2}{r}^{2}+1560\,{e}^{4}l{r}^{6}+80\,{e}^{4}lM{r}^{5} \big) \nonumber \\
	& & ~~~
	+ a^4 \big( 250\,e{r}^{6}-340\,eM{r}^{5}+2100\,e{l}^{2}{r}^{4}-4200\,e{l}^{2}M{r}^{3}-500\,e{l}^{4}{r}^{2}-5000\,e{l}^{4}Mr+280\,{e}^{3}{r}^{6}-1160\,{e}^{3}M{r}^{5}\nonumber \\
	& & ~~~~~~~~
	+280\,{e}^{3}{M}^{2}{r}^{4}-4500\,{e}^{3}{l}^{2}{r}^{4}-5200\,{e}^{3}{l}^{2}M{r}^{3}+2000\,{e}^{3}{l}^{2}{M}^{2}{r}^{2}-530\,{e}^{5}{r}^{6}-60\,{e}^{5}M{r}^{5} \big) \nonumber \\
	& & ~~~
	+ a^5 \big( -150\,l{r}^{4}-500\,{l}^{3}{r}^{2}-2250\,{e}^{2}l{r}^{4}+4200\,{e}^{2}lM{r}^{3}+2000\,{e}^{2}{l}^{3}{r}^{2}+10000\,{e}^{2}{l}^{3}Mr+3400\,{e}^{4}l{r}^{4}\nonumber \\
	& & ~~~~~~~~
	+3300\,{e}^{4}lM{r}^{3}-1000\,{e}^{4}l{M}^{2}{r}^{2} \big) \nonumber \\
	& & ~~~
	+ a^6 \big( 150\,e{r}^{4}+1500\,e{l}^{2}{r}^{2}+800\,{e}^{3}{r}^{4}-1400\,{e}^{3}M{r}^{3}-3000\,{e}^{3}{l}^{2}{r}^{2}-10000\,{e}^{3}{l}^{2}Mr-950\,{e}^{5}{r}^{4}\nonumber \\
	& & ~~~~~~~~
	-800\,{e}^{5}M{r}^{3}+200\,{e}^{5}{M}^{2}{r}^{2} \big) \nonumber \\
	& & ~~~
	+ a^7 \big( -1500\,{e}^{2}l{r}^{2}+2000\,{e}^{4}l{r}^{2}+5000\,{e}^{4}lMr \big) \nonumber \\
	& & ~~~
	+ a^8 \big( 500\,{e}^{3}{r}^{2}-500\,{e}^{5}{r}^{2}-1000\,{e}^{5}Mr \big)
	\Bigg] {\Delta \tau}^{4} \nonumber \\
	& &
	-\frac {3 q^2 M^2}{11200\left( {r}^{2}-2\,Mr+{a}^{2} \right) {r}^{16}}  \sqrt{e^2-1 - 2 V_{\rm{eff}}} \Bigg[
	\big( -36\,e{r}^{9}+81\,eM{r}^{8}-120\,e{l}^{2}{r}^{7}+264\,e{l}^{2}M{r}^{6}\nonumber \\
	& & ~~~~~~~~
	+180\,e{l}^{4}{r}^{5}-390\,e{l}^{4}M{r}^{4}+16\,{e}^{3}{r}^{9}+240\,{e}^{3}{l}^{2}{r}^{7} \big) \nonumber \\
	& & ~~~
	+ \big( 240\,l{r}^{7}-648\,lM{r}^{6}+270\,l{M}^{2}{r}^{5}+720\,{l}^{3}{r}^{5}-2040\,{l}^{3}M{r}^{4}+1056\,{l}^{3}{M}^{2}{r}^{3}-360\,{l}^{5}M{r}^{2}+780\,{l}^{5}{M}^{2}r\nonumber \\
	& & ~~~~~~~~
	+240\,{e}^{2}l{r}^{7}-816\,{e}^{2}lM{r}^{6}-2160\,{e}^{2}{l}^{3}{r}^{5}+1080\,{e}^{2}{l}^{3}M{r}^{4}-480\,{e}^{4}l{r}^{7} \big) a \nonumber \\
	& & ~~~
	+ \big( -456\,e{r}^{7}+921\,eM{r}^{6}-270\,e{M}^{2}{r}^{5}-3360\,e{l}^{2}{r}^{5}+8784\,e{l}^{2}M{r}^{4}-3168\,e{l}^{2}{M}^{2}{r}^{3}+2280\,e{l}^{4}{r}^{3}\nonumber \\
	& & ~~~~~~~~
	+4290\,e{l}^{4}M{r}^{2}-3900\,e{l}^{4}{M}^{2}r-104\,{e}^{3}{r}^{7}+552\,{e}^{3}M{r}^{6}+5640\,{e}^{3}{l}^{2}{r}^{5}-900\,{e}^{3}{l}^{2}M{r}^{4}+240\,{e}^{5}{r}^{7} \big) {a}^{2} \nonumber \\
	& & ~~~
	+ \big( 960\,l{r}^{5}-1368\,lM{r}^{4}+3520\,{l}^{3}{r}^{3}-6120\,{l}^{3}M{r}^{2}-4200\,{l}^{5}M+4560\,{e}^{2}l{r}^{5}-11448\,{e}^{2}lM{r}^{4}\nonumber \\
	& & ~~~~~~~~
	+3168\,{e}^{2}l{M}^{2}{r}^{3}-10560\,{e}^{2}{l}^{3}{r}^{3}-13560\,{e}^{2}{l}^{3}M{r}^{2}+7800\,{e}^{2}{l}^{3}{M}^{2}r-5520\,{e}^{4}l{r}^{5}+120\,{e}^{4}lM{r}^{4} \big) {a}^{3} \nonumber \\
	& & ~~~
	+ \big( -1140\,e{r}^{5}+1368\,eM{r}^{4}-11640\,e{l}^{2}{r}^{3}+18360\,e{l}^{2}M{r}^{2}+2100\,e{l}^{4}r+21000\,e{l}^{4}M-1920\,{e}^{3}{r}^{5}\nonumber \\
	& & ~~~~~~~~
	+4704\,{e}^{3}M{r}^{4}-1056\,{e}^{3}{M}^{2}{r}^{3}+18000\,{e}^{3}{l}^{2}{r}^{3}+18540\,{e}^{3}{l}^{2}M{r}^{2}-7800\,{e}^{3}{l}^{2}{M}^{2}r+1860\,{e}^{5}{r}^{5}\nonumber \\
	& & ~~~~~~~~
	+90\,{e}^{5}M{r}^{4} \big) {a}^{4} \nonumber \\
	& & ~~~
	+ \big( 720\,l{r}^{3}+2800\,{l}^{3}r+12720\,{e}^{2}l{r}^{3}-18360\,{e}^{2}lM{r}^{2}-8400\,{e}^{2}{l}^{3}r-42000\,{e}^{2}{l}^{3}M-13440\,{e}^{4}l{r}^{3}\nonumber \\
	& & ~~~~~~~~
	-11760\,{e}^{4}lM{r}^{2}+3900\,{e}^{4}l{M}^{2}r \big) {a}^{5} \nonumber \\
	& & ~~~
	+ \big( -720\,e{r}^{3}-8400\,e{l}^{2}r-4600\,{e}^{3}{r}^{3}+6120\,{e}^{3}M{r}^{2}+12600\,{e}^{3}{l}^{2}r+42000\,{e}^{3}{l}^{2}M+3720\,{e}^{5}{r}^{3}\nonumber \\
	& & ~~~~~~~~
	+2850\,{e}^{5}M{r}^{2}-780\,{e}^{5}{M}^{2}r \big) {a}^{6} \nonumber \\
	& & ~~~
	+ \big( 8400\,{e}^{2}lr-8400\,{e}^{4}lr-21000\,{e}^{4}lM \big) {a}^{7} \nonumber \\
	& & ~~~
	+ \big( -2800\,{e}^{3}r+2100\,{e}^{5}r+4200\,{e}^{5}M \big) {a}^{8}
	\Bigg] {\Delta \tau}^{5} + O \left( \Delta \tau ^6 \right)\\
\frac{dm}{d\tau} &=& \frac {3 q^2 M^2}{11200{r}^{15}} \Bigg[
	\left( -5\,{r}^{7}+10\,M{r}^{6}-40\,{l}^{2}{r}^{5}+80\,{l}^{2}M{r}^{4}-50\,{l}^{4}{r}^{3}+100\,{l}^{4}M{r}^{2}-20\,{e}^{2}{r}^{7}-80\,{e}^{2}{l}^{2}{r}^{5} \right)\nonumber \\
	& & ~~~
	+ a \left( 240\,el{r}^{5}-160\,elM{r}^{4}+600\,e{l}^{3}{r}^{3}-400\,e{l}^{3}M{r}^{2}+160\,{e}^{3}l{r}^{5} \right) \nonumber \\
	& & ~~~
	+ a^2 \left( -20\,{r}^{5}-300\,{l}^{2}{r}^{3}-500\,{l}^{4}r-200\,{e}^{2}{r}^{5}+80\,{e}^{2}M{r}^{4}-1500\,{e}^{2}{l}^{2}{r}^{3}+600\,{e}^{2}{l}^{2}M{r}^{2}-80\,{e}^{4}{r}^{5} \right)\nonumber \\
	& & ~~~
	+ a^3 \left( 600\,el{r}^{3}+2000\,e{l}^{3}r+1400\,{e}^{3}l{r}^{3}-400\,{e}^{3}lM{r}^{2} \right) \nonumber \\
	& & ~~~
	+ a^4 \left( -300\,{e}^{2}{r}^{3}-3000\,{e}^{2}{l}^{2}r-450\,{e}^{4}{r}^{3}+100\,{e}^{4}M{r}^{2} \right) \nonumber \\
	& & ~~~
	+ a^5 \left(2000\,{e}^{3}rl \right) - a^6 \left( 500 \,{e}^{4}r \right) \Bigg] {\Delta \tau}^{4} \nonumber \\
	& &
	+ \frac {3 q^2 M^2}{11200{r}^{15}} \sqrt{e^2-1 - 2 V_{\rm{eff}}} \Bigg[
	\big( 12\,{r}^{6}-27\,M{r}^{5}+120\,{l}^{2}{r}^{4}-264\,{l}^{2}M{r}^{3}+180\,{l}^{4}{r}^{2}-390\,{l}^{4}Mr\nonumber \\
	& & ~~~~~~~~
	+48\,{e}^{2}{r}^{6}+240\,{e}^{2}{l}^{2}{r}^{4} \big)\nonumber \\
	& & ~~~
	+ a \big( -720\,el{r}^{4}+528\,elM{r}^{3}-2160\,e{l}^{3}{r}^{2}+1560\,e{l}^{3}Mr-480\,{e}^{3}l{r}^{4} \big) \nonumber \\
	& & ~~~
	+ a^2  \big(60\,{r}^{4}+1080\,{l}^{2}{r}^{2}+2100\,{l}^{4}+600\,{e}^{2}{r}^{4}-264\,{e}^{2}M{r}^{3}+5400\,{e}^{2}{l}^{2}{r}^{2}-2340\,{e}^{2}{l}^{2}Mr+240\,{e}^{4}{r}^{4} \big) \nonumber \\
	& & ~~~
	+ a^3 \big( -2160\,el{r}^{2}-8400\,e{l}^{3}-5040\,{e}^{3}l{r}^{2}+1560\,{e}^{3}lMr \big)\nonumber \\
	& & ~~~
	+ a^4 \big( 1080\,{e}^{2}{r}^{2}+12600\,{e}^{2}{l}^{2}+1620\,{e}^{4}{r}^{2}-390\,{e}^{4}Mr \big) \nonumber \\
	& & ~~~
	- a^5 \big( 8400\,{e}^{3}l\big) + a^6 \big( 2100\,{e}^{4} \big)
	\Bigg] {\Delta \tau}^{5} + O \left( \Delta \tau ^6 \right)
\end{eqnarray}
\end{subequations}

It is straightforward to see that these results exactly match Eqs. (\ref{eq:ma-r-schw-el}) - (\ref{eq:dmdtau-schw-el}) in the limit $a \rightarrow 0$.

\subsubsection{Release from instantaneous rest}
\label{sec:kerr-equ-rest}
We now consider the motion of a particle which is released from rest relative to an observer at spatial infinity. This is analogous to the case of radial infall from rest in Schwarzschild (Section \ref{sec:rad-rest}) and we proceed with the calculation in exactly the same way. Imposing the initial conditions $u^r(r=r_0)=0$ and $u^\phi(r=r_0)=0$, Eqs. (\ref{eq:kerr-ur}) and (\ref{eq:kerr-uphi}) can be solved for the constants of motion:

\begin{subequations}
\begin{eqnarray}
\label{eq:kerr-e}
 e &=& \sqrt{\frac{r_0 -2 M}{r_0}}\\
\label{eq:kerr-l}
 l &=& -\frac{2Ma}{r_0}\sqrt{\frac{r_0}{r_0-2M}}
\end{eqnarray}
\end{subequations}

We then substitute these into Eq. (\ref{eq:kerr-ur}) and integrate to get an approximate expression for $r_0$ in terms of $r$:

\begin{equation}
 r_0 = r + \left(M r_0^2 - \left(l^2 - a^2 \left(e^2-1\right)\right)r_0 + 3M\left(l-ae\right)^2\right) \frac{\Delta \tau^2}{2r_0^4}
\end{equation}

This, along with the equations for $e$ and $l$ then allow us to express the 4-velocity as:
\begin{subequations}
\begin{eqnarray}
 u^r &=& -\left(\frac{r^2-2rM+a^2}{r^3\left(r-2M\right)}\right) M \Delta \tau\\
 u^\theta &=& 0\\
 u^\phi &=& \frac{1}{\Delta}\left[ -\left(\frac{r-2M}{r}\right)\frac{2Ma}{r}\sqrt{\frac{r}{r-2M}} + \frac{2Ma}{r}\sqrt{\frac{r-2 M}{r}}\right]\\
u^t &=& \frac{1}{\Delta} \left[\left(r^2 + a^2 + \frac{2Ma^2}{r}\right)\sqrt{\frac{r -2 M}{r}} + \frac{4M^2a^2}{r^2}\sqrt{\frac{r}{r-2M}}\right] 
\end{eqnarray}
\end{subequations}

We then substitute these values along with the components of the Hadamard/DeWitt coefficients into Eq. (\ref{eq:SimpForce}) and project orthogonal and tangent to the 4-velocity to get the equations of motion:

\begin{subequations}
\begin{eqnarray}
ma^r &=& -{\frac {3 {q}^{2}{M}^{2}}{11200 \left( {r}^{2}-2\,Mr+{a}^{2} \right) ^{3} \left( 2\,M-r \right) ^{3}{r}^{23}}}\, \Bigg[
	- \left( 39\,M-20\,r \right) \left( 2\,M-r \right) ^{7}{r}^{15} \nonumber \\
	& & ~~~
	-4 a^2 \, \left( 252\,{M}^{3}-160\,r{M}^{2}-162\,{r}^{2}M+95\,{r}^{3} \right)  \left( 2\,M-r \right) ^{6}{r}^{12} \nonumber \\
	& & ~~~
	- 12 a^4 \, \left( 848\,{M}^{5}-640\,r{M}^{4}-1404\,{r}^{2}{M}^{3}+960\,{r}^{3}{M}^{2}+291\,{r}^{4}M-185\,{r}^{5} \right)  \left( 2\,M-r \right) ^{5}{r}^{9} \nonumber \\
	& & ~~~
	-4 a^6 \, \big( 12288\,{M}^{7}-10240\,r{M}^{6}-42720\,{r}^{2}{M}^{5}+32640\,{r}^{3}{M}^{4}+20844\,{r}^{4}{M}^{3}-14880\,{r}^{5}{M}^{2}-2292\,{r}^{6}M\nonumber \\
	& & ~~~~~~~~
	+1545\,{r}^{7} \big)  \left( 2\,M-r \right) ^{4}{r}^{6} \nonumber \\
	& & ~~~
	- a^8 \big( 98304\,{M}^{9}-81920\,r{M}^{8}-811008\,{r}^{2}{M}^{7}+655360\,{r}^{3}{M}^{6}+758592\,{r}^{4}{M}^{5}-583680\,{r}^{5}{M}^{4}\nonumber \\
	& & ~~~~~~~~
	-187920\,{r}^{6}{M}^{3}+138240\,{r}^{7}{M}^{2}+13347\,{r}^{8}M-9420\,{r}^{9} \big)  \left( 2\,M-r \right) ^{3}{r}^{3} \nonumber \\
	& & ~~~
	+12 a^{10} \, \left( 8\,{M}^{2}-{r}^{2} \right)  \big( 15872\,{M}^{7}-12800\,r{M}^{6}-30272\,{r}^{2}{M}^{5}+24000\,{r}^{3}{M}^{4}+10704\,{r}^{4}{M}^{3}-8200\,{r}^{5}{M}^{2}\nonumber \\
	& & ~~~~~~~~
	-921\,{r}^{6}M+675\,{r}^{7} \big) \left( 2\,M-r \right) ^{2}{r}^{2} \nonumber \\
	& & ~~~
	-10 a^{12} \, \left( 2\,M-r \right) \left( 7296\,{M}^{5}-5760\,r{M}^{4}-4704\,{r}^{2}{M}^{3}+3680\,{r}^{3}{M}^{2}+489\,{r}^{4}M-370\,{r}^{5} \right)  \left( 8\,{M}^{2}-{r}^{2} \right) ^{2}r\nonumber \\
	& & ~~~
	+100 a^{14} \, \left( 9\,M-7\,r \right)  \left( 8\,{M}^{2}-{r}^{2} \right) ^{4}
	 \Bigg] {\Delta \tau}^{5} + O\left(\Delta \tau^6 \right) \\
ma^\theta &=& O\left(\Delta \tau^6 \right) \\
ma^\phi &=&{\frac {3{q}^{2}{M}^{2}a}{112 \left( -2\,Mr+{r}^{2}+{a}^{2} \right) ^{3} \left( 2\,M-r \right) ^{3/2}{r}^{{37/2}} }}\, \Bigg[
	{r}^{8} \left( 2\,M-r \right) ^{4}\nonumber \\
	& & ~~~
	+a^2 {r}^{5} \left( 16\,{M}^{2}-7\,{r}^{2} \right)  \left( 2\,M-r \right) ^{3}\nonumber \\
	& & ~~~
	+16\,a^4 {r}^{2} \left( 4\,{M}^{4}-6\,{M}^{2}{r}^{2}+{r}^{4} \right) \left( 2\,M-r \right) ^{2}\nonumber \\
	& & ~~~
	-5\,a^6 r \left( 2\,M-r \right)  \left( -3\,{r}^{2}+8\,{M}^{2} \right)  \left( 8\,{M}^{2}-{r}^{2} \right)\nonumber \\
	& & ~~~
	+5\, a^8 \left( 8\,{M}^{2}-{r}^{2} \right) ^{2}
	\Bigg] {\Delta \tau}^{4}+ O\left(\Delta \tau^6 \right) \\
ma^t &=&{\frac {3{q}^{2}{M}^{3}{a}^{2}}{56\left( {r}^{2}-2\,Mr+{a}^{2} \right) ^{5} \left( 2\,M-r \right) ^{5/2}{r}^{{43/2}}}}\, 
	\left( 2\,M{r}^{3}-{r}^{4}+ \left( 8\,{M}^{2}-{r}^{2} \right) {a}^{2} \right)  \Bigg[
	-{r}^{12} \left( 2\,M-r \right) ^{7}\nonumber \\
	& & ~~~
	-a^2 {r}^{9}\left( 40\,{M}^{3}-16\,r{M}^{2}-30\,{r}^{2}M+9\,{r}^{3} \right) \left( 2\,M-r \right) ^{5}\nonumber \\
	& & ~~~
	-a^4 {r}^{6} \left( 256\,{M}^{5}-64\,r{M}^{4}-560\,{r}^{2}{M}^{3}+128\,{r}^{3}{M}^{2}+138\,{r}^{4}M-31\,{r}^{5} \right)  \left( 2\,M-r \right) ^{4}\nonumber \\
	& & ~~~
	-2\,a^6 {r}^{3} \left( 256\,{M}^{7}-1664\,{M}^{5}{r}^{2}+224\,{M}^{4}{r}^{3}+1100\,{M}^{3}{r}^{4}-184\,{M}^{2}{r}^{5}-147\,M{r}^{6}+27\,{r}^{7} \right)  \left( 2\,M-r \right) ^{3}\nonumber \\
	& & ~~~
	+a^8 {r}^{2} \left( 6144\,{M}^{7}-10752\,{M}^{5}{r}^{2}+1024\,{M}^{4}{r}^{3}+3680\,{M}^{3}{r}^{4}-496\,{M}^{2}{r}^{5}-324\,M{r}^{6}+51\,{r}^{7} \right)  \left( 2\,M-r \right) ^{2}\nonumber \\
	& & ~~~
	- 5\,a^{10}\,r \left( 2\,M-r \right)  \left( -5\,{r}^{5}+36\,{r}^{4}M+24\,{r}^{3}{M}^{2}-272\,{r}^{2}{M}^{3}+384\,{M}^{5} \right)  \left( 8\,{M}^{2}-{r}^{2} \right)\nonumber \\
	& & ~~~
	+5 a^{12}\, \left( 32\,{M}^{3}-8\,{r}^{2}M+{r}^{3} \right)  \left( 8\,{M}^{2}-{r}^{2} \right) ^{2}
	 \Bigg] {\Delta \tau}^{4} + O\left(\Delta \tau^6 \right) \\
\frac{dm}{d\tau} &=& -{\frac {3 {q}^{2}{M}^{2}}{448{r}^{20} \left( {r}^{2}-2\,Mr+{a}^{2} \right) ^{4} \left( 2\,M-r \right)^{2}}}\, \Bigg[
	{r}^{15} \left( r-2\,M \right) ^{7}\nonumber \\
	& & ~~~
	+16\, a^2 {r}^{12} \left( {r}^{2}-2\,{M}^{2} \right)  \left( r-2\,M \right) ^{6}\nonumber \\
	& & ~~~
	+12\, a^4 {r}^{9} \left( 7\,{r}^{4}-40\,{r}^{2}{M}^{2}+32\,{M}^{4} \right)  \left( r-2\,M \right) ^{5}\nonumber \\
	& & ~~~
	+8\, a^6 {r}^{6} \left( 27\,{r}^{6}-276\,{M}^{2}{r}^{4}+672\,{M}^{4}{r}^{2}-256\,{M}^{6} \right)  \left( r-2\,M \right) ^{4}\nonumber \\
	& & ~~~
	+ a^8 {r}^{3} \left( {r}^{2}-8\,{M}^{2} \right)  \left( -512\,{M}^{6}+3264\,{M}^{4}{r}^{2}-2232\,{M}^{2}{r}^{4}+309\,{r}^{6} \right)  \left( r-2\,M \right) ^{3}\nonumber \\
	& & ~~~
	+12\, a^{10} {r}^{2} \left( 21\,{r}^{4}-96\,{r}^{2}{M}^{2}+64\,{M}^{4} \right) \left( r-2\,M \right) ^{2} \left( {r}^{2}-8\,{M}^{2} \right) ^{2}\nonumber \\
	& & ~~~
	+10\, a^{12} r \left( r-2\,M \right)  \left( -24\,{M}^{2}+11\,{r}^{2} \right)  \left( {r}^{2}-8\,{M}^{2} \right) ^{3}\nonumber \\
	& & ~~~
	+20\, a^{14} \left( {r}^{2}-8\,{M}^{2} \right) ^{4}
	\Bigg] {\Delta \tau}^{4}+ O\left(\Delta \tau^6 \right) 
\end{eqnarray}
\end{subequations}

Again, it is straightforward to see that these results exactly match Eqs. (\ref{eq:RadialRestQLSF}) - (\ref{eq:RadialRestdmdtau}) in the limit $a \rightarrow 0$.

\section{Coordinate Calculation}
\label{sec:coord}
Using a Hadamard-WKB expansion, Anderson and Hu \cite{Anderson:2003qa}\footnote{It is important to use the corrected expressions for the $v_{ijk}$ as given in the erratum to this paper.}  calculated the tail part of the retarded Green's function in Schwarzschild spacetime in terms of the Schwarzschild coordinates up to $O\left( \left( x-x' \right)^6 \right)$ in the separation of the points. They then used their results as a basis for the calculation of the entire self-force on a particle held at rest until a time $t = 0$, at which point it is released and allowed to fall freely \cite{Anderson:2005ds}. Due to the spherical symmetry of the Schwarzschild spacetime, the resulting motion is radially inwards.

The results of Ref. \cite{Anderson:2003qa} were also used by Anderson and Wiseman \cite{Anderson:2005gb} to calculate the self-force on a particle in Schwarzschild spacetime under two other circumstances: a static particle and a particle following a circular geodesic. Unfortunately, although the technique they describe is valid, there are some errors in the results. For this reason, and in order to extract the radially infalling part from the results of Ref. \cite{Anderson:2005ds}, we reproduce those calculations here with corrections. We also go one step further and calculate the equations of motion from the self-force. Although the static particle result also contains errors, we will not consider it here as it corresponds to non-geodesic motion, which we have considered, but leave for later work to avoid confusion.

We follow the method prescribed by Anderson and Wiseman \cite{Anderson:2005gb} and begin with the same expression we had for the covariant calculation:
\begin{equation}
\label{eq:QLSFInt-coord}
f^{a}_{\rm QL} = -q^2 \int_{\tau - \Delta \tau}^{\tau} \nabla^{a} v \xxp d\tau '
\end{equation}

Next, instead of using a covariant expansion for $v \xxp$ as we did previously, we now use the coordinate expansion given in Ref. \cite{Anderson:2003qa}\footnote{There are several different conventions in the literature for the Hadamard form of the retarded Green's function. For the present work, we use a convention consistent with Ref. \cite{Decanini:2005gt} throughout. However, the convention of Ref. \cite{Anderson:2003qa} differs by a factor of $\frac{1}{4\pi}$ in the coefficient of $v \xxp$. Fortunately, our convention for the scalar charge, $q$ also differs by a factor of $\sqrt{4\pi}$, with the result that our covariant $V \xxp$ is exactly equal to the coordinate $v \xxp$ of Anderson and Hu \cite{Anderson:2003qa}. Furthermore, Ref. \cite{Anderson:2003qa} originally indicated a factor of $\frac{1}{8\pi}$ in the coefficient of $v \xxp$. It has subsequently been confirmed with the authors that this should be $\frac{1}{4\pi}$.}:
\begin{equation}
\label{eq:WKBGreen}
v\xxp = \sum_{i,j,k=0}^{\infty} v_{ijk} \left( t - t' \right)^{2i} \left( cos \left( \gamma \right) - 1 \right)^j \left(r - r'\right)^k 
\end{equation}
The coordinate $\gamma$ is the angle on the 2-sphere between $x$ and $x'$. For the spherically symmetric Schwarzschild spacetime, we can set $\theta=\frac{\pi}{2}$ without loss of generality, so that $\gamma=\phi$. We then take the partial derivative with respect to the coordinates $r$, $\phi$, and $t$ and convert from those coordinates to proper time $\tau$. Next, we integrate over $\tau$ as we did previously in the covariant calculation and project orthogonal to and along the 4-velocity to obtain a final expression for the quasi-local contribution to the equations of motion. This coordinate transformation is dependent on the particle's path so it is in this way that the motion of the particle affects the self-force and equations of motion. We now work through this calculation for two of the particle paths investigated previously in the covariant calculation.

\subsection{Circular Geodesic}
\label{sec:coord-circ}
For a circular geodesic, the geodesic equations can be integrated to give
\begin{equation}
	r-r'=0,~ ~ ~\theta - \theta'=0,~ ~ ~\phi - \phi'=\frac{1}{r}\sqrt{\frac{M}{r-3M}}({\tau}-{\tau}'),~ ~ ~
	t-t'=\sqrt{\frac{r}{r-3M}}({\tau}-{\tau}')
\end{equation}
We now substitute Eq. (\ref{eq:WKBGreen}) into (\ref{eq:QLSFInt-coord}), take the partial derivative, then use the above to express the coordinate separations in terms of proper time separations. Finally, we do the easy integration over $\tau'$ and project orthogonal and parallel to the 4-velocity to obtain an expression for the quasi-local contribution to the equations of motion for a particle following a circular geodesic in Schwarzschild spacetime. The radial component of the quasi-local mass times 4-acceleration is
\begin{equation}
\label{eq:rCoordCirc}
ma_{\rm QL}^{r} = ma_{\rm QL}^{r} \left[ 5 \right] {\Delta}{\tau}^5 + ma_{\rm QL}^{r} \left[ 7 \right] {\Delta}{\tau}^7 + O\left({\Delta}{\tau}^9\right)
\end{equation}
where
\begin{subequations}
\begin{eqnarray}
ma_{\rm QL}^{r} \left[ 5 \right] &=& -\frac{3 q^2 M^2 \left( r-2M \right) \left( 20 r^3 - 81 M r^2 + 54 M^2 r + 53 M^3 \right)}{11200 \left( r-3M \right) ^2 r^{11}} \\
ma_{\rm QL}^{r} \left[ 7 \right] &=& -\frac{q^2 M^2 \left( r-2M \right) }{188160 \left( r-3M \right)^3 r^{14} } \times \nonumber \\
	& & ~~~
	\left( 560 r^5 - 6020 r^4 M + 21465 r^3 M^2 - 26084 r^2 M^3 - 5281 rM^4 + 21510 M^5 \right)
\end{eqnarray}
\end{subequations}

The $\theta$ component of the quasi-local mass times 4-acceleration $0$ as would be expected.

The $\phi$ component of the quasi-local mass times 4-acceleration is:
\begin{equation}
\label{eq:phiCoordCirc}
ma_{\rm QL}^{\phi} = ma_{\rm QL}^{\phi} \left[ 4 \right] {\Delta}{\tau}^4 + ma_{\rm QL}^{\phi} \left[ 6 \right] {\Delta}{\tau}^6 + O\left({\Delta}{\tau}^8\right)
\end{equation}
where
\begin{subequations}
\begin{eqnarray}
ma_{\rm QL}^{\phi} \left[ 4 \right] &=& \frac{3 q^2 M^2 \left( r-2M \right)^2 \left( 7r-8M \right)}{1120 \left( r-3M \right) ^2 r^{10}} \sqrt{\frac{M}{r-3M}}\\
ma_{\rm QL}^{\phi} \left[ 6 \right] &=& \frac{q^2 M^2 \left( r-2M \right) }{13440 r^{13} \left( r-3M \right) ^3} \sqrt{\frac{M}{r-3M}} \left( 126 r^4 - 1129 r^3 M + 3447 M^2 r^2 -4193 M^3 r  + 1623 M^4 \right)
\end{eqnarray}
\end{subequations}

The $t$ component of the quasi-local mass times 4-acceleration is:
\begin{equation}
\label{eq:tCoordCirc}
ma_{\rm QL}^{t} = ma_{\rm QL}^{t} \left[ 4 \right] {\Delta}{\tau}^4 + ma_{\rm QL}^{t} \left[ 6 \right] {\Delta}{\tau}^6 + O\left({\Delta}{\tau}^8\right)
\end{equation}
where
\begin{subequations}
\begin{eqnarray}
ma_{\rm QL}^{t} \left[ 4 \right] &=& \frac{3 q^2 M^3 \left( r-2M \right) \left( 7r-8M \right)}{1120 \left( r-3M \right)^2 r^{9}} \sqrt{\frac{r}{r-3M}}\\
ma_{\rm QL}^{t} \left[ 6 \right] &=& \frac{q^2 M^3}{13440 \left( r-3M \right)^3 r^{12}}  \sqrt{\frac{r}{r-3M}} \left( 126 r^4 - 1129 r^3 M + 3447 r^2 M^2 - 4193 r M^3 + 1623 M^4 \right)
\end{eqnarray}
\end{subequations}

Finally, the quasi-local contribution to the mass change is:
\begin{equation}
\frac{dm}{d\tau} = \frac{dm}{d\tau} \left[ 4 \right] {\Delta}{\tau}^4 + \frac{dm}{d\tau} \left[ 6 \right] {\Delta}{\tau}^6 + O\left({\Delta}{\tau}^8\right)
\end{equation}
where
\begin{subequations}
\begin{eqnarray}
\frac{dm}{d\tau} \left[ 4 \right] &=& \frac { 3 q^2 M^2 \left( r-2M \right) \left( 13M^2 + 2Mr-  5r^2 \right) }{2240 \left( r-3M \right) ^2 r^9} \\
\label{eq:dmdtauCoordCirc}
\frac{dm}{d\tau} \left[ 6 \right] &=& \frac { q^2 M^2  \left( r-2M\right) \left( 651\,{M}^{4}-344\,{M}^{3}r-261\,{M}^{2}{r}^{2}+189\,M{r}^{3}-28\,{r}^{4} \right)}{ 6720 \left( r-3M \right)^3 r^{12}}
\end{eqnarray}
\end{subequations}

Direct comparison between these results for the equations of motion and those of Ref. \cite{Anderson:2005gb} for the self-force are not immediately possible. However, it is straightforward to project the results of Ref. \cite{Anderson:2005gb} orthogonal to and tangent to the 4-velocity. In doing so, we find that both sets of results agree in their general structure, but differ in the numerical coefficients involved. We suspect a minor algebraic slip may have led to the errors in the results of Ref. \cite{Anderson:2005gb}.

\subsection{Radial Geodesic: Infall from rest}
\label{sec:coord-radial}
The case of a particle following a radial geodesic has been investigated by Anderson, Eftekharzadeh and Hu \cite{Anderson:2005ds}. They consider a particle held at rest until a time $t=0$. The particle is then allowed to freely fall radially inward. In such a case they have shown that, provided the particle falls for a sufficiently short amount of time, it is possible to obtain an expression for the entire self-force, rather than just the quasi-local part. Unfortunately, direct comparison with our result is not immediately possible as their results give the full self-force, rather than just the quasi-local contribution from the radial infall phase. However, as it is straightforward to calculate quasi-local contribution using the method they prescribe, we have done so and found agreement with our results once the corrections to Ref. \cite{Anderson:2003qa} mentioned earlier in this section are taken into account.

As we are only interested in the leading order behavior for comparison to our own results, we take a slightly more straightforward approach to the calculation. From the arguments in Section \ref{sec:rad-rest}, it is clear that

\begin{equation}
\label{eq:coord-tau-radrest}
	r-r'\approx-\frac{M}{2 r^2}\left(\tau^2 - \tau'^2\right),~ ~ ~\theta - \theta'=0,~ ~ ~\phi - \phi'=0 ,~ ~ ~
	t-t'\approx\sqrt{\frac{r}{r-2M}}({\tau}-{\tau}')
\end{equation}

As in the circular geodesic case, we now substitute Eq. (\ref{eq:WKBGreen}) into (\ref{eq:QLSFInt-coord}), take the partial derivative, then use Eq. (\ref{eq:coord-tau-radrest}) to express the coordinate separations in terms of proper time separations. Finally, we do the integration and project orthogonal and parallel to the 4-velocity to obtain an expression for the quasi-local contribution to the equations of motion for a particle starting at rest at $r=r_0, \tau = 0$ and subsequently falling radially inwards for a proper time $\Delta\tau$ to a radius $r$

\begin{subequations}
\begin{eqnarray}
\label{eq:RadialRest-coord}
ma_{\rm QL}^{r} &=&
	- \frac {3 q^2 M^2}{11200 r^{10}}   \left( \frac{r-2M}{r} \right) \left( 20r-39M \right) {\Delta \tau}^5 + O \left( \Delta \tau ^6 \right)\\
ma_{\rm QL}^{\theta} &=& 0 + O \left( \Delta \tau ^6 \right) \\
ma_{\rm QL}^{\phi} &=& 0 + O \left( \Delta \tau ^6 \right) \\
ma_{\rm QL}^{t} &=& 0 + O \left( \Delta \tau ^6 \right) \\
\label{eq:RadialRestdmdtau-coord}\frac{dm}{d\tau} &=&
	- \frac {3 q^2 M^2}{448 r^8} \left(\frac{r - 2M}{r} \right) {\Delta \tau}^4 + O \left( \Delta \tau ^6 \right) 
\end{eqnarray}
\end{subequations}

This is in exact agreement with our covariant results in Eqs. (\ref{eq:RadialRestQLSF}) - (\ref{eq:RadialRestdmdtau}).
 
\subsection{Local Truncation Error}
As our results differ from those of Ref. \cite{Anderson:2005gb}, it is desirable to also update the plots of the relative error given there. We estimate the local fractional truncation error, as Anderson and Wiseman have done, by the ratio between the highest order term in the expansion ( $O \left( \Delta \tau ^n \right)$, say) and the sum of all the terms up to that order,

\begin{equation}
\epsilon \equiv \frac{f^a_{\rm QL}\left[ n \right]}{\sum_{i=0}^{n} f^a_{\rm QL}\left[ i \right]}
\end{equation}

\begin{figure}[htb!]
\centering%
\includegraphics[width=7cm]{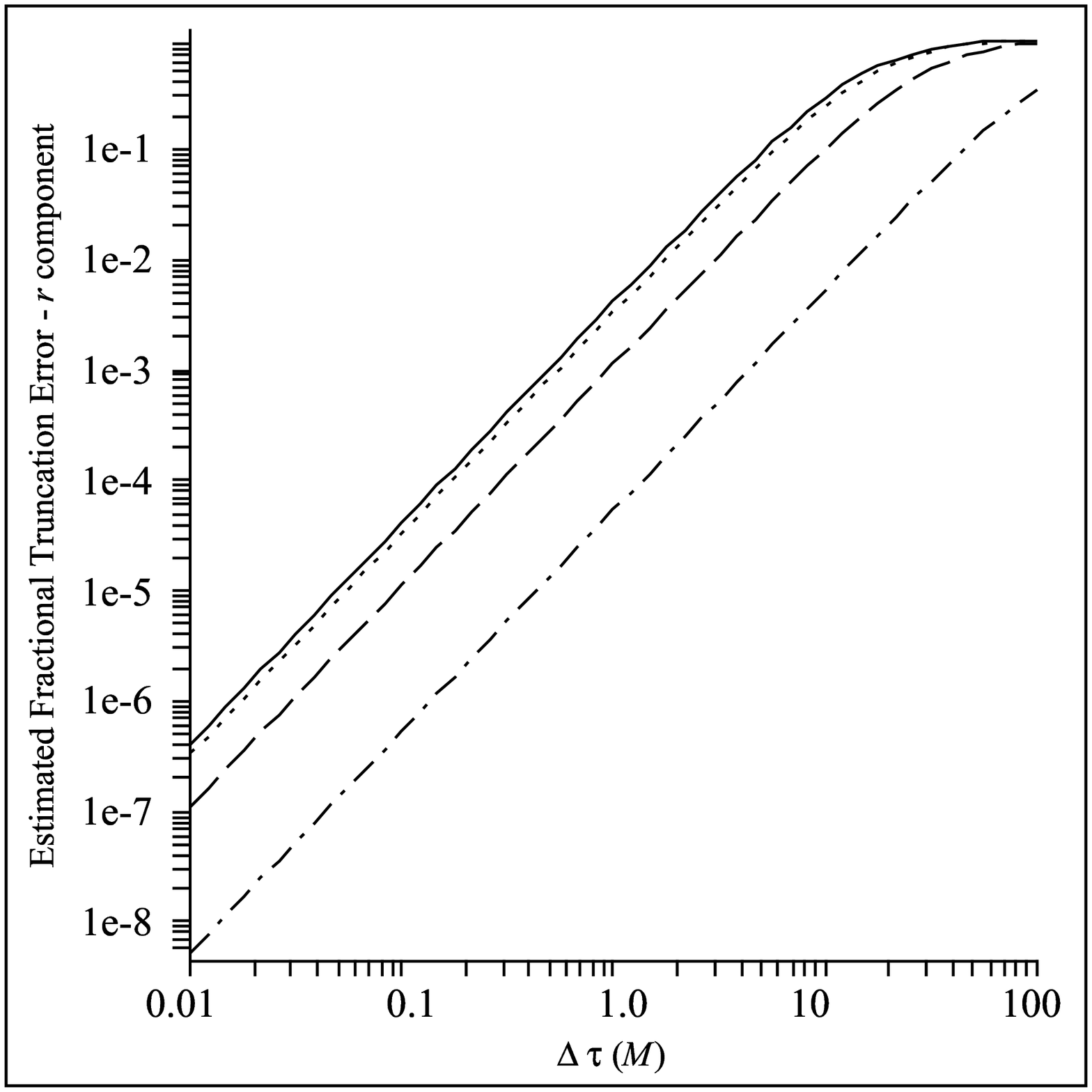}
\includegraphics[width=7cm]{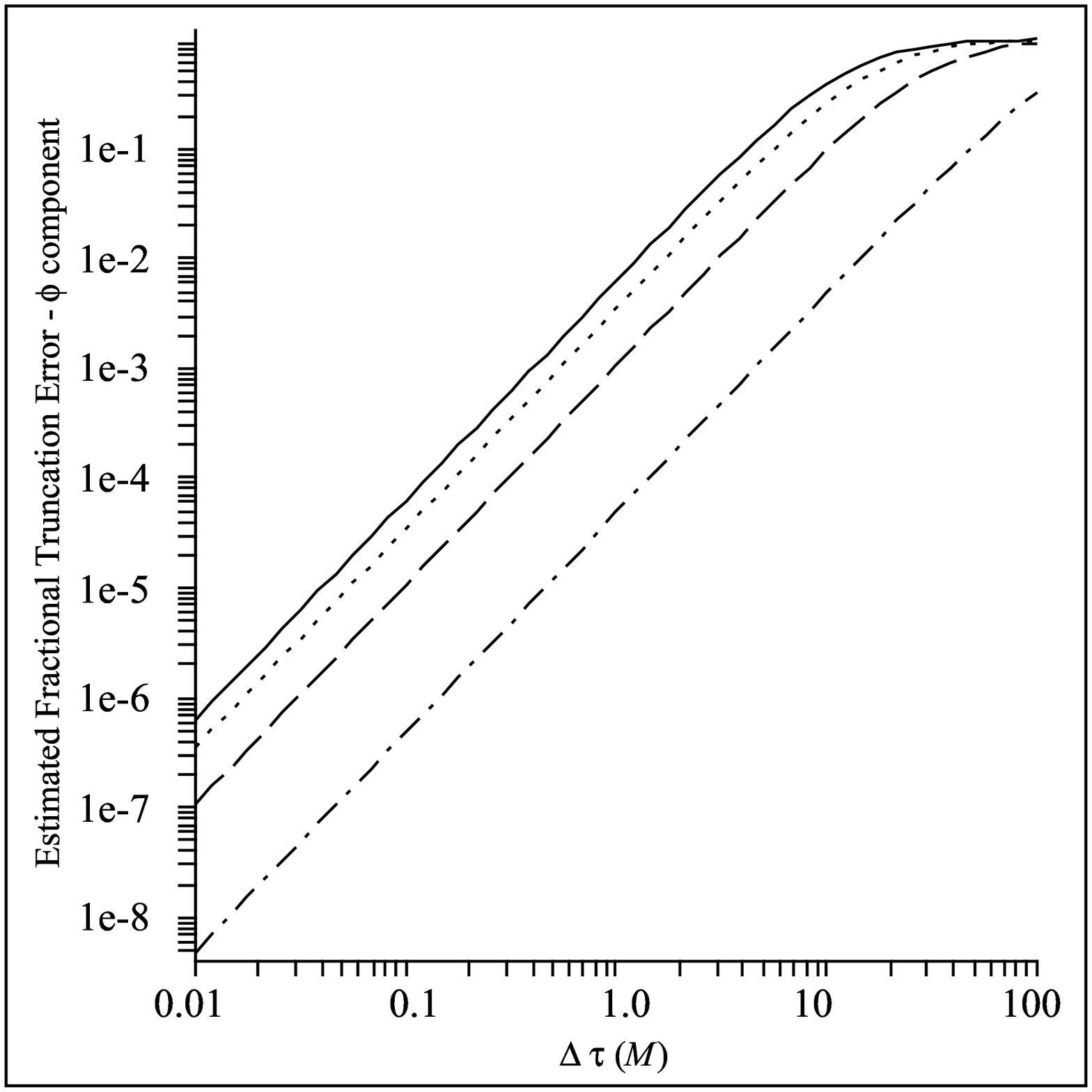}
\includegraphics[width=7cm]{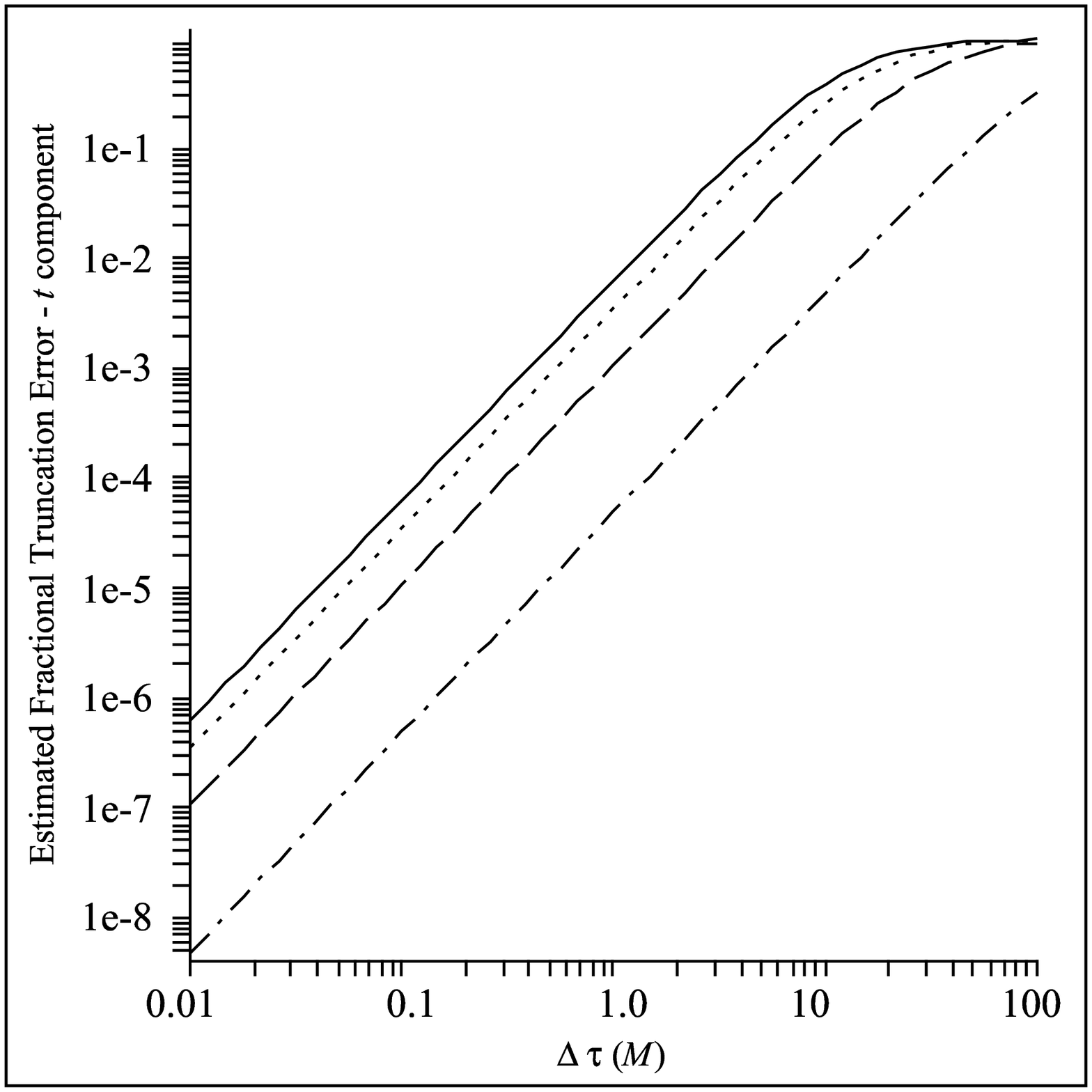}
\includegraphics[width=7cm]{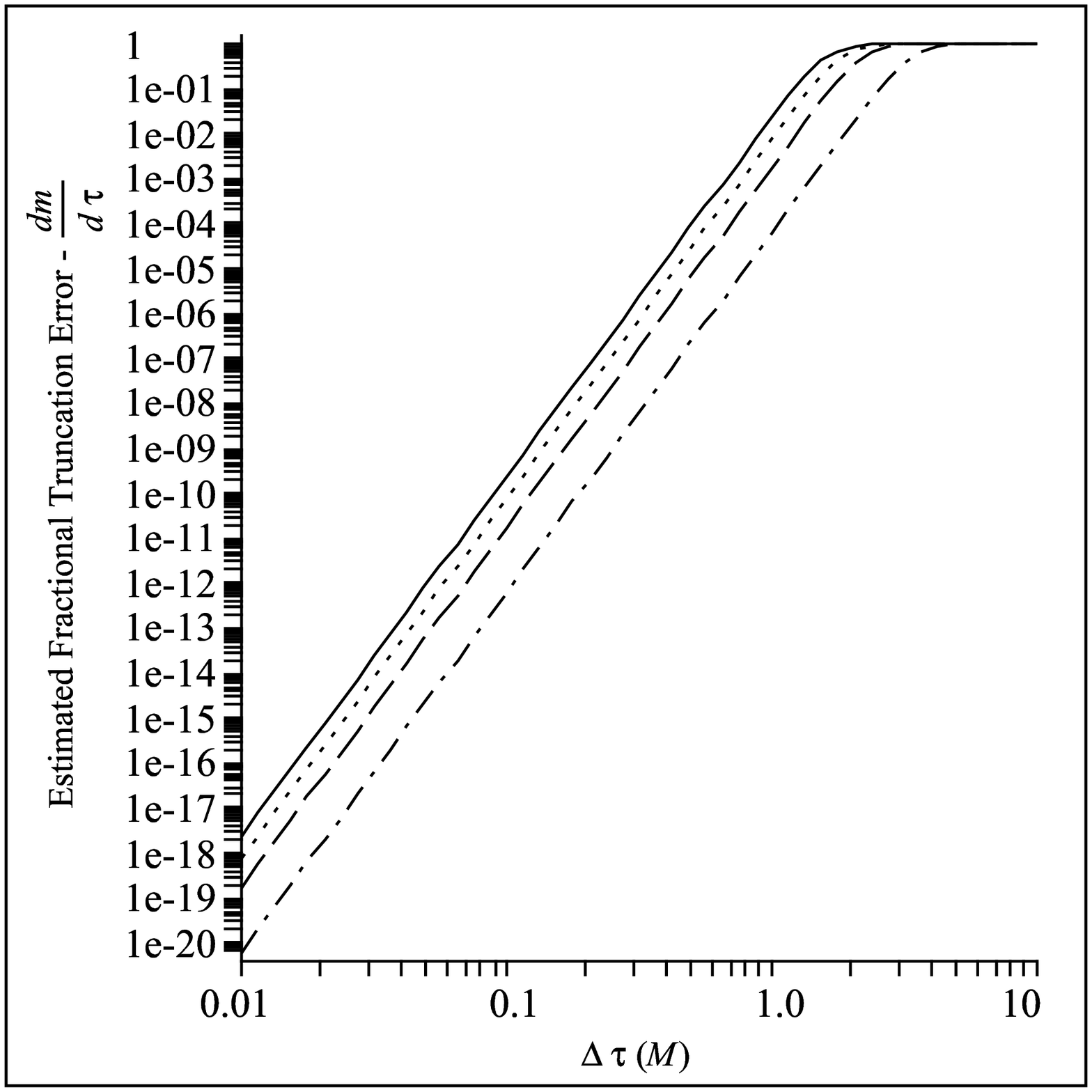}
\includegraphics[width=8cm]{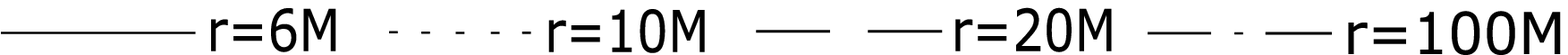}
\caption{Local truncation error in the equations of motion for a particle following a circular geodesic in Schwarzschild.}
\label{fig:error}
\end{figure}

For the circular geodesic motion investigated in the present work, we plot in Fig. \ref{fig:error} the fractional local truncation error for the equations of motion as a function of $\Delta \tau$ for $r=6M$, $r=10M$, $r=20M$ and $r=100M$, as was done previously for the self-force in Ref. \cite{Anderson:2005gb}.

\section{Conclusion}
The self-force is determined by the entire past history of the particle, while the quasi-local contribution is the portion of the self-force resulting from the particle's recent past. In this paper, we have calculated the quasi-local contribution to the self-force and to the equations of motion for a scalar charge moving in a curved background spacetime. We have determined these quantities up to fifth order in $\Delta \tau$, the proper time distance backwards along the particle's path to which the quasi-local contribution applies. In addition to this result, which applies to a general spacetime, we also present the same result for vacuum spacetimes and show how the expressions simplify significantly.

We then looked at some specific examples of spacetimes and particle motions. We looked at the Schwarzschild and Kerr spacetimes and at various geodesic motions. This allowed us to compare our results to existing work and to verify the validity of our approach. It also allowed us to find some errors in some of the existing work.

While Anderson, Eftekharzadeh and Hu have successfully calculated the quasi-local contribution to the self-force to a much higher order (in $\Delta \tau$) than we have done here, their results apply only to a specific choice of spacetime and to a particular particle motion. On the other hand, the covariant method we describe only provides terms at the leading two orders for massless fields in vacuum spacetimes, but does so for a general vacuum spacetime and general (geodesic) particle motion. In this sense, the two methods are seen as complimentary, each proving as a useful check on the others results.

We have chosen to restrict the present work to geodesic motion. This was desirable primarily in order to avoid confusion. Non-geodesic motion adds some extra complexity to the covariant calculation due to there being two different proper times: the particle proper time and the geodesic proper time. We have made significant progress in dealing with this issue and intend to present our solution to the problem at a later date.

Obviously, the scalar case as presented here is of less immediate physical interest than the gravitational case. The biggest obstacle to achieving results in the gravitational case will be to calculate the relevant DeWitt/Hadamard coefficients. While this has previously been done for vacuum spacetimes up to $O(\sigma^{3/2})$ by Anderson, Flanagan and Ottewill \cite{Anderson:2004eg}, it is desirable to take the calculation to higher order to allow the matching point $\Delta \tau$ to lie further into the particle's past. It would also be desirable, but not necessary, to obtain expressions for a general spacetime, rather than restricting ourselves to only vacuum spacetimes. Achieving these goals would be possible using a method similar to that of Folacci and D\'{e}canini \cite{Decanini:2005gt}, for example. However, it is probable that the expressions would be even longer for spin-2 (gravitational) than for spin-0 (scalar) fields. Some hope lies in the fact that we are most interested calculating $V (x,x')$ for massless fields in 4-dimensional vacuum spacetimes. There is also hope that work by Fulling, et al. \cite{Fulling:1992vm} and more recently by D\'{e}canini and Folacci \cite{Decanini:2007preprint} to establish a linearly independent basis for the Riemann polynomials will simplify the calculation somewhat. Once $V (x,x')$ has been calculated to sufficiently high order, it will be straightforward to apply a method analogous the the one presented here to calculate the self-force.

\section{Acknowledgments}
BW is supported by the Irish Research Council for Science, Engineering and Technology: funded by the National Development Plan. We would like to thank  Paul Anderson and Ardeshir Eftekharzadeh for the considerable time and effort they gave to resolving some issues that arose during this work. We would also like to thank Warren Anderson and Antoine Folacci whose ideas and suggestions were invaluable.  Also our appreciation to the many others from the Capra 10 meeting for interesting and useful conversations. Finally, we would like to thank Marc Casals and Sam Dolan for their continuous feedback and discussions.
\appendix
\section{Calculating covariant Hadamard/De-Witt coefficients to $5^{th}$ order in the separation of the points}
\label{sec:cov-5}
Folacci and D\'{e}canini \cite{Decanini:2005gt} have calculated the DeWitt and Hadamard coefficients of the Feynman propagator in a general spacetime up to $O\left( \sigma^2 \right)$ in the separation of the points (i.e. fourth order in the geodesic separation of the points). In particular, we are interested here in the Hadamard coefficients for the function $V\left( x, x' \right)$, which appears in the Hadamard form of the Feynman propagator.

This may be extended to $O\left( \sigma^{5/2} \right)$ relatively easily by taking into account the symmetry $V\xxp = V\left( x',x \right)$ of the Green's function. Repeatedly taking symmetrized covariant derivatives of both sides of this equation and taking coincidence limits, we arrive at an expressions for the odd order terms in terms of all the lower (even) order terms \cite{Brown:1986tj}:
\begin{equation}
\label{eq:OddEven}
\sum_{r=0}^{n} \binom{n}{r} V_{(a_1 a_2 \dots a_r ; a_{r+1} \dots a_n )} = (-1)^n V_{a_1 a_2 \dots a_n}
\end{equation}

Recursively applying this identity to eliminate all odd order terms, we get expressions for the coefficients of the terms of order $\sigma^\frac{5}{2}$:

\begin{subequations}
\begin{eqnarray}
\label{eq:cov-5}
v_{abcde} &=& \frac{1}{2} v_{;(a b c d e)} - \frac{5}{2} v_{(a b ; c d e)} + \frac{5}{2} v_{(a b c d ; e)} \\
v_{abc} &=& -\frac{1}{4} v_{;(a b c)} + \frac{3}{2} v_{(a b; c)} \\
v_{a} &=& \frac{1}{2} v_{;a}
\end{eqnarray}
\end{subequations}

These expressions are equally valid for any symmetric bi-tensor, so we may use it for both the calculation of higher order $v_{n a_1 \dots a_p}$ as used in (\ref{eq:V}) and higher order $v_{a_1 \dots a_p}$ as used in (\ref{eq:Vt}).

\section{Vacuum expressions for the Hadamard/DeWitt Coefficients}
\label{sec:DeWittVacuum}
The expressions given in Ref. \cite{Decanini:2005gt} are unnecessarily long and unwieldy for the purposes of the present work. We are currently only interested in massless fields in vacuum spacetimes ($m_{\rm field}=0$, $R_{\alpha \beta}=0$), so the majority of the terms vanish and we are left with much more manageable expressions for the $v_{n a_1 \dots a_p}$. They are:

\begin{subequations}
\begin{eqnarray}
\label{eq:V0}
v_0 &=& 0\\
v_{0\, a} &=& 0\\
v_{0\, ab} &=& - \frac{1}{720} g_{a b} I\\
v_{0\, abc} &=& - \frac{1}{480} g_{(a b} I_{;c)}\\
v_{0\, abcd} &=& - \frac{2}{525} C^{\rho}_{\phantom{\rho} (a | \sigma | b} \Box C^{\sigma}_{\phantom{\sigma} c | \rho | d)}
	- \frac{2}{105}C^{\rho \sigma \tau}_{\phantom{\rho \sigma \tau} (a} C_{| \rho \sigma \tau | b ; c d)}
	- \frac{1}{280} C^{\rho \phantom{ (a | \sigma | b} ;\tau}_{\phantom{\rho} (a | \sigma | b} C^{\sigma}_{\phantom{\sigma} c | \rho | d) ;\tau}
	\nonumber \\
	& &
	- \frac{1}{56} C^{\rho \sigma \tau}_{\phantom{\rho \sigma \tau} (a ;b} C_{| \rho \sigma \tau | c ; d)} 
	- \frac{2}{1575} C^{\rho \sigma \tau \kappa} C_{\rho (a | \tau | b} C_{| \sigma | c | \kappa | d)}
	- \frac{2}{525} C^{\rho \kappa \tau}_{\phantom{\rho \kappa \tau} (a} C^{\phantom{| \rho \tau | } \sigma}_{| \rho \tau | \phantom{\sigma} b} C_{| \sigma | c | \kappa | d)}
	\nonumber \\
	& &
	- \frac{8}{1575} C^{\rho \kappa \tau}_{\phantom{\rho \kappa \tau} (a} C^{\phantom{| \rho | } \sigma}_{| \rho | \phantom{ \sigma } | \tau | b} C_{| \sigma | c | \kappa | d)}
	- \frac{4}{1575} C^{\rho \tau \kappa}_{\phantom{\rho \tau \kappa} (a} C^{\phantom{| \rho \tau | } \sigma}_{| \rho \tau | \phantom{\sigma} b} C_{| \sigma | c | \kappa | d)}\\
\label{eq:V1}
v_1 &=& \frac{1}{720} I \\
v_{1\, a} &=& \frac{1}{1440} I_{;a} \\
v_{1\, ab} &=& - \frac{1}{1400} C^{\rho \sigma \tau \kappa} C_{\rho \sigma \tau (a ; b) \kappa}
	+ \frac{1}{1575} C^{\rho \sigma \tau}_{\phantom{\rho \sigma \tau} a} \Box C_{\rho \sigma \tau b}
	+ \frac{29}{25200} C^{\rho \sigma \tau \kappa} C_{\rho \sigma \tau \kappa ; (a b)}
	\nonumber \\
	& &
	+ \frac{1}{1680} C^{\rho \sigma \tau}_{\phantom{\rho \sigma \tau} a ;\kappa} C^{\phantom{\rho \sigma \tau b} ;\kappa}_{\rho \sigma \tau b}
	+\frac{1}{1344} C^{\rho \sigma \tau \kappa}_{\phantom{\rho \sigma \tau \kappa} ;a} C_{\rho \sigma \tau \kappa ;b}
	+ \frac{1}{756} C^{\rho \kappa \sigma \lambda} C^{\tau}_{\phantom{\tau} \rho \sigma a} C_{\tau \kappa \lambda b}
	\nonumber \\
	& &
	- \frac{1}{1800} C^{\rho \kappa \sigma \lambda} C_{\rho \sigma \tau a} C^{\phantom{\kappa \lambda} \tau}_{\kappa \lambda \phantom{\tau} b}
	+ \frac{19}{18900} C^{\rho \sigma \kappa \lambda} C_{\rho \sigma \tau a} C^{\phantom{\kappa \lambda} \tau}_{\kappa \lambda \phantom{\tau} b}\\
\label{eq:V2}
v_2 &=& - \frac{1}{6720} C_{\rho \sigma \tau \kappa} \Box C^{\rho \sigma \tau \kappa}
	- \frac{1}{8960} C_{\rho \sigma \tau \kappa ; \lambda} C^{\rho \sigma \tau \kappa ; \lambda}
	\nonumber \\
	& &
	- \frac{1}{9072} C^{\rho \kappa \sigma \lambda} C_{\rho \alpha \sigma \beta} C^{\phantom{\kappa} \alpha \phantom{\lambda} \beta}_{\kappa \phantom{\alpha} \lambda}
	- \frac{11}{18440} C^{\rho \sigma \kappa \lambda} C_{\rho \sigma \alpha \beta} C^{\phantom{\kappa \lambda} \alpha \beta}_{\kappa \lambda}
\end{eqnarray}
\end{subequations}
where $C_{a b c d}$ is the Weyl tensor and $I = C_{\alpha \beta \gamma \delta}C^{\alpha \beta \gamma \delta}$.

Moreover, we have chosen to work with an expression for $V(x,x')$ as given in (\ref{eq:Vt}), rather than the expression used in Ref. \cite{Decanini:2005gt} and Eq. (\ref{eq:V}). The two sets of coefficients may be related by using 

\begin{equation}
2 \sigma = \sigma^{;\alpha} \sigma_{;\alpha}
\end{equation}
to rewrite (\ref{eq:V}) in terms of a series expansion in powers of $\sigma^{;\alpha}$. We can then equate, power by power, the coefficients of powers of $\sigma^{;\alpha}$ to the corresponding terms in (\ref{eq:Vt}) to give:

\begin{subequations}
\begin{eqnarray}
v 		&=& v_0\\
v_a 		&=& v_{0\, a}\\
v_{ab} 		&=& v_{0\, ab} + v_{1}g_{ab}\\
v_{abc}		&=& v_{0\, abc} + 3 v_{1\, (a}g_{bc)}\\
v_{abcd}	&=& v_{0\, abcd} + 6 v_{1\, (ab}g_{cd)} + 6 v_2 g_{(ab}g_{cd)}\\
v_{abcde}	&=& \frac{1}{2} v_{;(a b c d e)} - \frac{5}{2} v_{(a b ; c d e)} + \frac{5}{2} v_{(a b c d ; e)} 
\end{eqnarray}
\end{subequations}

Filling in the values for the $v_{n\, a_1 \dots a_p}$ we see that a large number of terms either cancel or combine to give:

\begin{subequations}
\begin{eqnarray}
v 		&=& 0\\
v_a 		&=& 0\\
v_{ab} 		&=& 0\\
v_{abc}		&=& 0\\
v_{abcd}	&=&-\frac{1}{280}C^{\rho \phantom{(a} \sigma}_{\phantom{\rho} (a \phantom{\sigma} b |;\alpha|}C^{\phantom{|\rho| c |\sigma| d)} ;\alpha}_{|\rho| c |\sigma| d)}
		   -\frac{2}{315}C^{\rho \sigma \tau \kappa}C_{\rho (a |\tau| b}C_{|\sigma| c |\kappa| d)}
		   +\frac{1}{105}C^{\rho \phantom{(a} \sigma}_{\phantom{\rho} (a \phantom{\sigma} b}C^{\tau \kappa}_{\phantom{\tau \kappa} |\rho| c}C_{|\tau \kappa \sigma| d)}\nonumber \\*
		& &+\frac{1}{840}C^{\rho \sigma \tau \kappa}C^{\phantom{\rho \sigma} \lambda}_{\rho \sigma \phantom{\lambda} (a}C_{|\tau \kappa \lambda| b}g_{c d)}
		   + \frac{1}{8960}\Box I g_{(a b}g_{c d)}
		   -\frac{1}{40320}C^{\rho \sigma \tau \kappa}C^{\phantom{\rho \sigma} u v}_{\rho \sigma}C_{\tau \kappa u v}g_{(a b}g_{c d)} \\
v_{abcde}	&=& \frac{5}{2} v_{(a b c d ;e )}
\end{eqnarray}
\end{subequations}

Here, we have used a slight modification of the basis of Riemann tensor polynomials suggested by Fulling, et al. \cite{Fulling:1992vm} in order to express the result in its simplest possible form. We have chosen an equivalent basis for Weyl tensor polynomials by ignoring all terms involving $R_{a b}$ and $R$ and replacing $R_{abcd}$ by $C_{abcd}$. We have also ignored all terms which vanish or are not independent under symmetrization of the free indices. Next, we have eliminated terms involving two covariant derivatives of a tensor in favor of terms involving two covariant derivatives of a scalar and single covariant derivatives of a tensor. We have done this as it is computationally faster and easier to calculate the covariant derivative of a scalar than of a tensor. Finally, we have used the identity \cite{Gunther}: 
\begin{equation}
 \nabla_{(r)} C^{\rho \sigma \tau}_{\phantom{\rho \sigma \tau} (a} \nabla_{(s)} C_{|\rho \sigma \tau | b)} = \frac{1}{4} g_{(a b)} \nabla_{(r)} C^{\rho \sigma \tau \kappa} \nabla_{(s)} C_{\rho \sigma \tau \kappa}
\end{equation}
(where $\nabla_{(r)}$ indicates $r$ covariant derivatives) to combine some of the remaining terms.

The motivation for working with $V(x,x')$ rather than the $V_n(x,x')$ is immediately apparent from the vast simplification that occurs.

\section{Calculating coordinate Hadamard/DeWitt coefficients to $7^{th}$ order in the separation of the points}
\label{sec:WKB-7}
In order to obtain expressions for the quasi-local equations of motion up to 7th order in the separation of the points, it is first necessary to extend the results of Ref. \cite{Anderson:2003qa} by one order. As is noted by Ref. \cite{Anderson:2005ds} and previously by Ref. \cite{Brown:1986tj}, this is a relatively easy calculation due to the symmetry in the Green's function:
\begin{equation}
v\xxp = v\left( x',x \right)
\end{equation}
We use expression (\ref{eq:WKBGreen}) for $v \xxp$ in the above to give: 
\begin{equation}
\sum_{i,j,k=0}^{\infty} v_{ijk} \left( r \right) \left( t - t' \right)^{2i} \left( \cos \left( \gamma \right) - 1 \right)^j \left(r - r'\right)^k =
	\sum_{i,j,k=0}^{\infty} v_{ijk} \left( r' \right)  \left( t' - t \right)^{2i} \left( \cos \left( \gamma ' \right) - 1 \right)^j \left(r' - r\right)^k \label{eq:SymGreen}
\end{equation}
Now, in a similar vein to Appendix \ref{sec:cov-5}, if we take a sufficient number of symmetrized derivatives of both sides and then take the coincidence limit, it is possible to express odd order coefficients in terms of derivatives of the lower order coefficients.

For the cases of relevance to the current work, the calculation is made even simpler by spherical symmetry. The coefficients $v_{ijk}$ are functions of $r$ and $M$ only (i.e. they have no dependence on $t$ or $\gamma$). Additionally, since $(t-t')^{2i}$ and $(\cos(\gamma)-1)^j$ are both even functions, they are invariant under interchange of primed and unprimed coordinates.  This means that we can express (\ref{eq:SymGreen}) in the simpler form:

\begin{equation}
\sum_{k=0}^{\infty} v_{ijk}\left( r \right) \left(r - r'\right)^k = \sum_{k=0}^{\infty} v_{ijk}\left( r' \right) \left(r' - r\right)^k
\end{equation}
Taking $r$-derivatives of both sides of this equation and taking the coincidence limit $r' \rightarrow r$ gives the result

\begin{equation}
v_{ij1} = - \frac{1}{2} v_{ij0,r}
\end{equation}

For the present work, only the coefficients $v_{301}$, $v_{211}$, $v_{121}$ and $v_{031}$ are needed. They are given here for convenience:

\begin{subequations}
\begin{eqnarray}
v_{301} &=& - \frac{1}{960}   \frac{M^2 \left( r -2M \right) ^2 \left( 598 r M^2 - 195 M r^2 + 20 r^3 - 585 M^3 \right) }{r^{16}} \\
v_{211} &=& \frac{1}{896}   \frac{M^2 \left( r -2M \right)  \left( 3352 r M^2 - 1099 M r^2 + 112 r^3 - 3240 M^3 \right) }{r^{13}} \\
v_{121} &=& - \frac{9}{1120} \frac{M^2 \left( 228 r M^2 - 91 M r^2 + 12 r^3 - 189 M^3 \right) }{r^{10}} \\
v_{031} &=& \frac{1}{3360} \frac{M^2 \left( -155 M r + 56 r^2 + 36 M^2 \right) }{r^7}
\end{eqnarray}
\end{subequations}





\end{document}